\documentclass[11pt]{article}

\usepackage[preprint]{acl}

\usepackage{times}
\usepackage{latexsym}

\usepackage[T1]{fontenc}
\usepackage[utf8]{inputenc}

\usepackage{microtype}

\usepackage{inconsolata}

\usepackage{graphicx}

\title{\emph{Who's important?---}\\\includegraphics[height=0.7em]{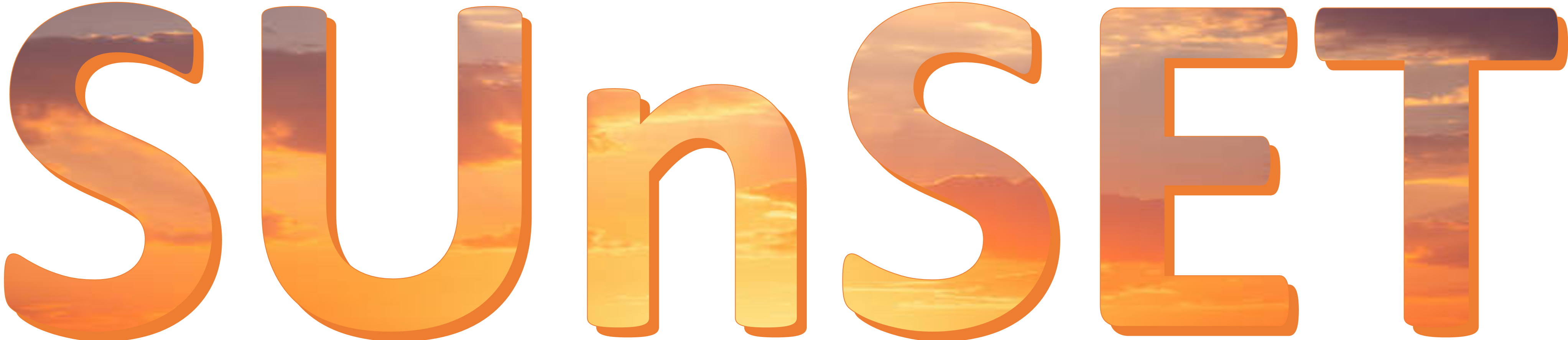}: Synergistic Understanding of Stakeholder, \\Events and Time for Timeline Generation}

\author{Tiviatis Sim$^{1,2}$, Yang Kaiwen$^{3}$, Shen Xin$^{3}$, Kenji Kawaguchi$^{1}$ \\
$^{1}$National University of Singapore, $^{2}$A*STAR Institute of High Performance Computing,\\
$^{3}$Celia Research Team, Norbert Wiener Research Center, Huawei International\\
\small{tiviatis@u.nus.edu}, \small{kaiwen\_yang@alumni.brown.edu}, \small{sxstar@zju.edu.cn}, \small{kenji@comp.nus.edu.sg}, 
}

\usepackage{multirow}
\usepackage{amsmath}
\usepackage{booktabs} %for table midrule etc.
\usepackage{tcolorbox} %for tcolorrbox
\usepackage{algorithm}
\usepackage{algorithmic}
\usepackage{changepage} % for adjustwidth*
\usepackage{subcaption} %for subfigure
\usepackage[table]{xcolor} % Enables row coloring in tables
\usepackage{amsthm} %proof/theorem stuff
\usepackage{amssymb} %for mathbb

\begin{document}
\maketitle
\begin{abstract}
As news reporting becomes increasingly global and decentralized online, tracking related events across multiple sources presents significant challenges. Existing news summarization methods typically utilizes Large Language Models and Graphical methods on article-based summaries. However, this is not effective since it only considers the textual content of similarly dated articles to understand the gist of the event. To counteract the lack of analysis on the parties involved, it is essential to come up with a novel framework to gauge the importance of stakeholders and the connection of related events through the relevant entities involved. Therefore, we present SUnSET: {\underline S}ynergistic {\underline U}{\underline n}derstanding of {\underline S}takeholder, {\underline E}vents and {\underline T}ime for the task of Timeline Summarization (TLS). We leverage powerful Large Language Models (LLMs) to build SET triplets and introduced the use of stakeholder-based ranking to construct a $Relevancy$ metric, which can be extended into general situations. Our experimental results outperform all prior baselines and emerged as the new State-of-the-Art, highlighting the impact of stakeholder information within news article. 
\end{abstract}

\section{Introduction}

The abundance of online news media in the advent of the information era has led to a growing volume of daily data production~\citep{Statista}, which poses significant challenges in efficiently identifying and understanding information concisely~\citep{Arnold_Goldschmitt_Rigotti_2023}. To promote efficacy and prevent information overloading, automated tasks such as Timeline Summarization (TLS)~\citep{2004paper, li-etal-2021-timeline, hu-etal-2024-moments} becomes especially critical in this era~\citep{Qorib_Hu_Ng_2025,wu-etal-2025-unfolding}. TLS is the generation of a summarized timeline of events, where multiple date-event pairs are sequentially listed to form a narrative for a particular topic. This process generally involves two essential requirements; (1) the events generated must be sequential and succinct (2) only relevant, fundamental and essential events should be included.
%%%%%%%%%%%%%%%%%%%%%%%%%%%%%%%%
\begin{figure}[t]
\centering
\includegraphics[width=0.55\textwidth, trim={0.3cm 0cm 0cm 0cm},clip]{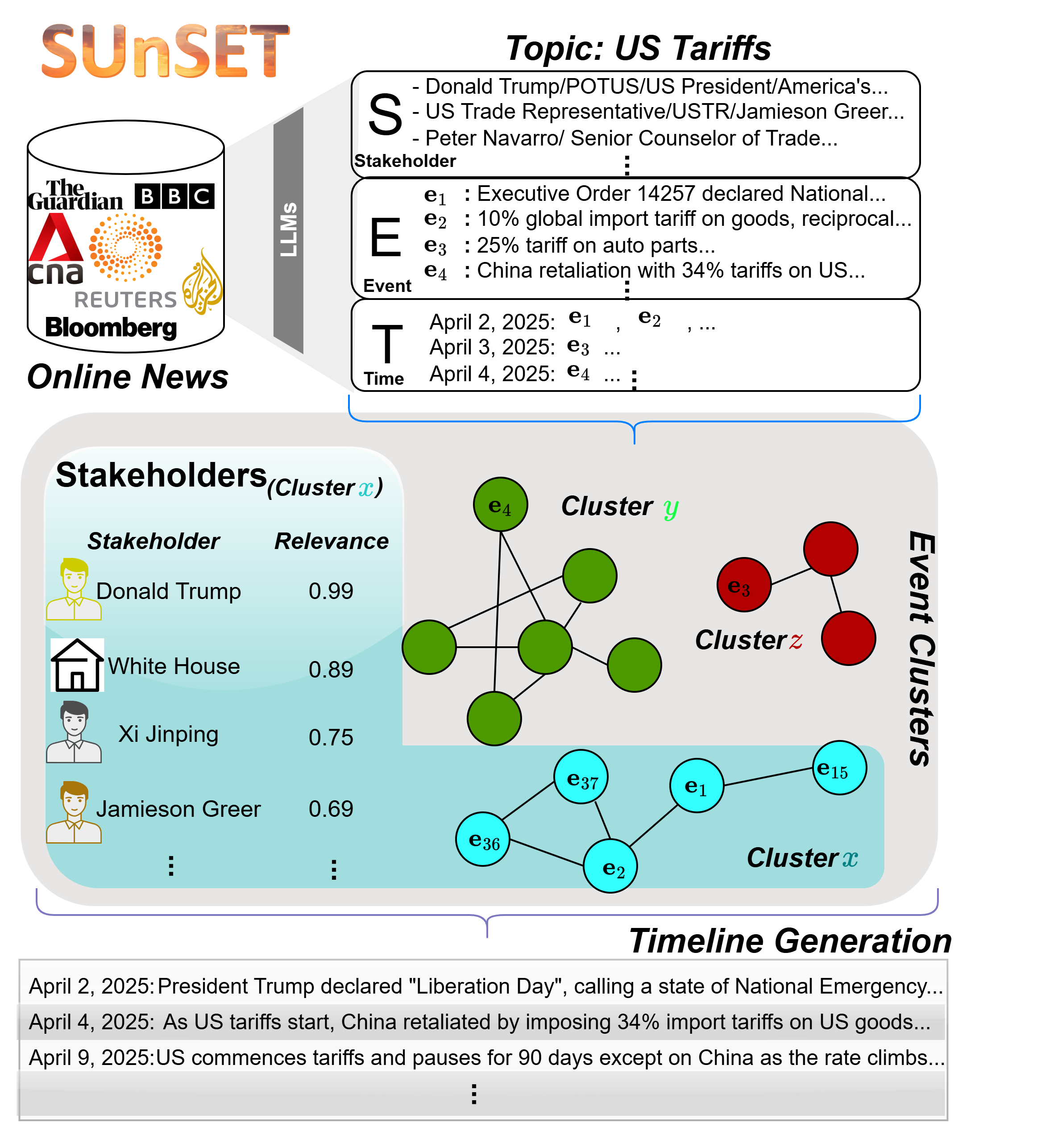}
% \vspace{-3mm}
\caption{\small{ 
An illustration of how SunSET generates a timeline for TLS through utilizing stakeholder information for relevance scoring.
}}
\vspace{-3mm}
\label{fig:formateval-teaser}
\end{figure}
%%%%%%%%%%%%%%%%%%%%%%%%%%%%%%%% 

Several papers have attempted to improve the brevity of the first requirement and the identification of the second through the use of powerful Large Language Models (LLMs). These methods typically yield good results~\citep{wang2023webnewstimelinegeneration,sojitra, hu-etal-2024-moments,Qorib_Hu_Ng_2025,wu-etal-2025-unfolding} due to the emergence of temporal understanding capabilities found within LLMs~\citep{xiong2024largelanguagemodelslearn}. Nevertheless, the utilization of LLMs in TLS is still relatively underexplored~\citep{hu-etal-2024-moments,sojitra} as compared to works on temporal reasoning such as temporal question answering tasks~\citep{piryani2025itshightimesurvey}.

Although previous works have utilized LLMs to aid temporal understanding within TLS frameworks, they typically focus on article summarization. Moreover, earlier approaches predating LLMs primarily addressed causal and temporal influences. Yet, stakeholders directly involved in events are crucial for identifying links between events and assessing their significance---an aspect largely overlooked in current TLS research. Incorporating stakeholder information can enhance both sequential curation and importance selection by grounding the timeline in real-world relevance.

%%%%%%%%%%%%%%%%%%%%%%%%%%%%%%%%

\begin{figure*}[t]
  \centering
  \includegraphics[width=1\textwidth, trim={0cm 0cm 0cm 0cm},clip]{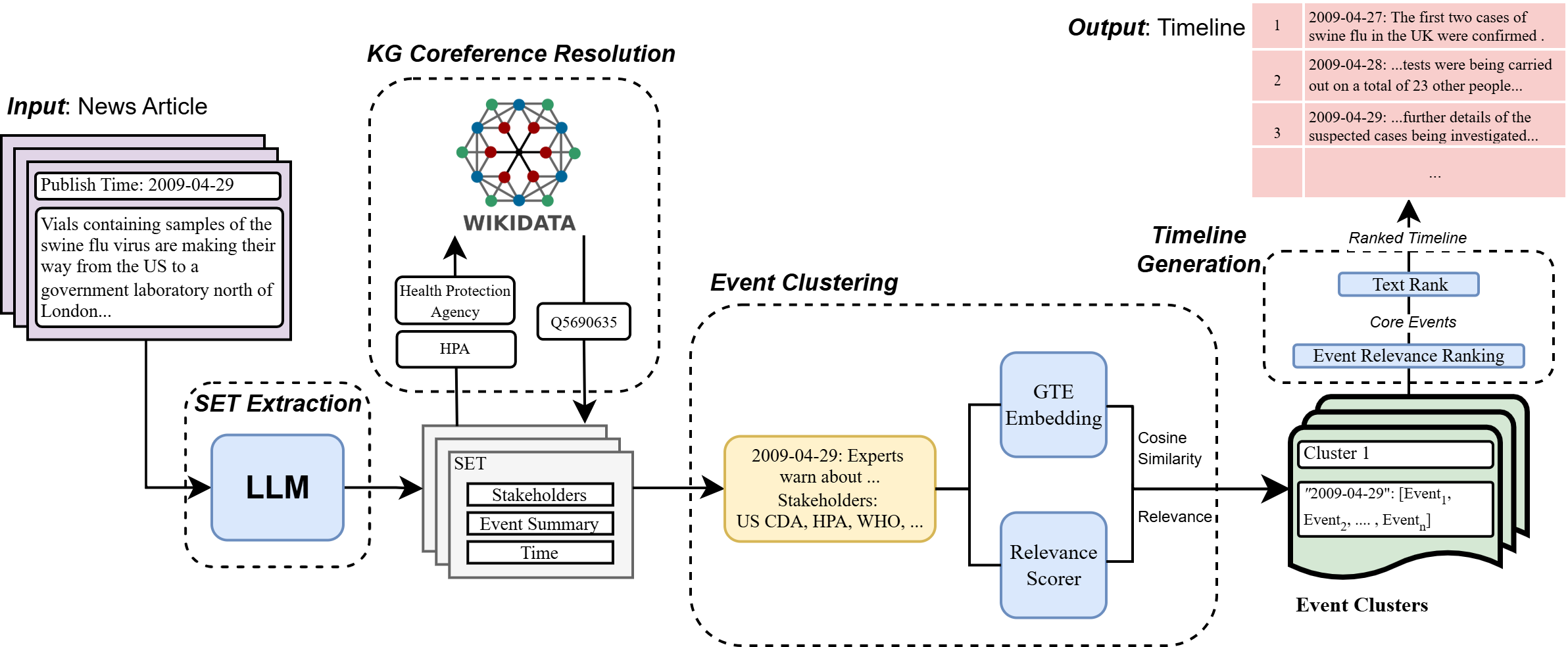}
  \caption{\small{ 
Full SUnSET Framework for TLS. News articles generate SETs and Stakeholders undergo Coreference Resolution. Subsequently, events are clustered through Cosine Similarity and Relevance. The clusters will then be ranked while TextRank extracts the narrative for final timeline creation. 
}}
\vspace{-3mm}
  \label{fig:framework}
\end{figure*}
%%%%%%%%%%%%%%%%%%%%%%%%%%%%%%%%
As such, we propose SUnSET: Synergistic Understanding of Stakeholder, Events and Time, for the task of automated TLS generation (Fig.~\ref{fig:formateval-teaser}). Unlike prior methods, SUnSET attempts to extract relevant information of various stakeholders mentioned within articles. Moreover, instead of summarizing articles immediately~\citep{Qorib_Hu_Ng_2025,hu-etal-2024-moments,someart}, SUnSET attempts to extract multiple relevant events mentioned within an article to enable a better representation of occurrences documented by a single news article. The extracted stakeholder and event information will subsequently be combined with its estimated timeframe to generate a Stakeholder-Event-Time triplet (SET). This will be utilized in various clustering steps of the timeline generation process to eventually generate a singular summarized timeline. SUnSET was tested against classic TLS datasets~\citep{t17,crisis} and was able to improve existing baselines to emerge as the new state of the art.

 Henceforth, the key contributions from our paper include the following:
\begin{enumerate}
    \item A novel framework is introduced to construct and apply SET for TLS.
    \item This is the first work to propose methods for extracting and utilizing stakeholder information within TLS.
    \item Multiple scoring mechanisms are developed, along with constraint-based proofs of stakeholder relevance applicable to general cases.
    \item Empirical evaluations across TLS datasets demonstrate improved performance, validating the benefits of SUnSET in TLS tasks.
\end{enumerate}

\section{Related Works} \label{ssec:related-work}

Recent TLS approaches can be broadly categorized into two main types: traditional methods and LLM-augmented methods; although variants of traditional clustering methods are still popular approaches when it comes to TLS task~\citep{gholipour-ghalandari-ifrim-2020-examining, chen2023followtimelinegeneratingabstractive}, the rise of emergent capabilities in generative LLMs has led to increasing efforts to leverage these tools for TLS tasks, often with diverse objectives. Although LLM applications are more commonly associated with temporal reasoning, their usage in TLS has grown significantly over the past three years. For instance, \citeauthor{wu-etal-2025-unfolding}'s paper highlighted the importance of temporal and causal relationships by incorporating LLMs into self-questioning strategies. On the other hand, works by \citeauthor{hu-etal-2024-moments} and \citeauthor{Qorib_Hu_Ng_2025} introduced the notion of LLMs functioning as pseudo-oracles, emulating crowdsourced event clustering via pairwise querying during the clustering process.

Clustering methods in TLS are typically implemented in two distinct ways~\citep{gholipour-ghalandari-ifrim-2020-examining}. The more conventional approach involves grouping key dates to generate graph-based rankings~\citep{steen-markert-2019-abstractive} where subsequent summaries will be formed by selecting the top candidate sentences derived from generated ranked lists of important dates. Alternatively, the other method involves `event' clustering~\citep{ribeiro-etal-2017-unsupervised}, where article-level summaries generated once from each article are aggregated based on textual similarity. Increasingly, LLMs are integrated into these processes, especially to enhance summarization. For example, \citeauthor{hu-etal-2024-moments} applied LLMs for article-level summarization prior to clustering, followed by a reclustering stage where the LLM validates cluster cohesion through pairwise comparisons of summary nodes. Meanwhile, \citeauthor{wu-etal-2025-unfolding} explored iterative prompting techniques to survey context by progressively self-refining queries with LLMs to retrieve more relevant content. Building upon these efforts, our paper advances TLS clustering with LLM augmentation, where we will expand on stakeholder-based heuristics to assess event relevance. 

\section{Methodology}
We introduce \textbf{SUnSET: Synergistic Understanding of Stakeholder, Events and Time} in this paper for the task of TLS. SUnSET aims to utilize stakeholder information while identifying multiple event occurrences within a single article for creating a graph representation of important events interconnected to each other. The general pipeline can be seen in Figure~\ref{fig:framework}, where the pipeline is split into three main sequences.

\subsection{SET Generation}
The first step of SUnSET requires the generation of Stakeholder, Event and Time to form a triplet (SET). This is done via utilizing a LLM to extract events mentioned within an article and their estimated date of occurrence. Therefore, every article $A$ may contain multiple events: $A\rightarrow \mathbf{e}_1,\mathbf{e}_2,\mathbf{e}_3,...$. After extracting the event and its estimated date ($t$), we call the same LLM to extract relevant stakeholders mentioned within the article with a maximum of 5 stakeholders per event: $\mathcal{S}_\mathbf{e}=(\varsigma_1,\varsigma_2,\varsigma_3,\varsigma_4,\varsigma_5)$. The term stakeholder used here strictly refers to an entity which is either a person or an organization. This results in a series of SET for a single article: $A\rightarrow((\mathcal{S}_{\mathbf{e}1},\mathbf{e}_1,t_{\mathbf{e}1}),(\mathcal{S}_{\mathbf{e}2},\mathbf{e}2,t_{\mathbf{e}2}),(\mathcal{S}_{\mathbf{e}3},\mathbf{e}3,t_{\mathbf{e}3}),...)$. The relevant prompts used can be found in Appendix~\ref{appn:prompts}. To minimize disruptions from hallucination, we used heuristic to ensure the generated date and event exists within the article and removed illogical dates. For events with lack of reference dates due to the lack of information within the article, the event uses the date of publication.

Subsequently, Coreference Resolution will be done for all of the extracted stakeholders due to difference in naming such as utilizing a title or a position than a name (E.g. President of the United States) or differences in naming (E.g. POTUS v.s. President of America). A knowledge graph (KG) was built from the wikidata~\citep{wikidata} database. This allows us to capture entities by origin-language forms and transliterations common within non-English dominant regions such as Semitic languages. The specifics of the KG building can be accessed in Appendix~\ref{appn:wikidata}.

\subsection{Event Clustering}

\begin{table}[h]
\centering
\resizebox{0.35\textwidth}{!}{

\begin{tabular}{ c c|c|c}
    & \multicolumn{3}{c}{\textbf{Rarity Across Topics}} \\
  \multirow{3}{*}{\rotatebox{90}{\textbf{Reoccurrence}}}
    &  & \textit{Rare} & \textit{Common} \\  \cline{2-4}
    & \rotatebox{90}{\textit{Low} } & Normal & Irrelevant \\  \cline{2-4}
    & \rotatebox{90}{\textit{High} } & Significant & Normal \\  
\end{tabular}}

\caption{Matrix showing Stakeholder frequency and relevance; Reoccurrence refers to repeated occurrence within the current topic, whereas rarity looks into frequency of a stakeholder across all topics.}
\label{tab:freq_table}
\end{table}
The next stage of the workflow involves the event clustering process. Typically, this involves using content found within similarly dated articles to generate clusters, where each article forms an event node. Instead of using the first five sentences in the body of
each article~\citep{gholipour-ghalandari-ifrim-2020-examining} or a pre-generated article summary~\citep{hu-etal-2024-moments,Qorib_Hu_Ng_2025} as the input, SUnSET uses every single event ($\mathbf{e}$) to generate a more accurate representation of an event's impact. This clustering process automatically reduces error brought forward by hallucination not caught in the previous step since hallucinations generate small event clusters which would be neglected due to their small event size. 

Additionally, unlike the aforementioned clustering methods which uses only date or cosine similarity for cluster creation, we came up with a new metric to gauge the importance of an event. This new metric incorporates stakeholder information; intuitively, the more the stakeholders occurs across all topics $D$, the less important they are. Similarly, if there is frequent reoccurrence of stakeholders within the same topic ($d\in D$), they are more significant (Tab.~\ref{tab:freq_table}). Therefore, SUnSET clustering introduces the use of Equation~\ref{eqn:relevance} to represent the relevance of stakeholders. This score penalizes common stakes across $D$ (Eqn.~\ref{eqn:P2}) while rewarding within-topic reoccurrence (Eqn.~\ref{eqn:R}).

\begin{equation}
 \label{eqn:relevance}
 Rel(\varsigma, d)= \beta (P(\varsigma,d)\cdot R(count(\varsigma_d)))
 \end{equation}
  \begin{equation}
 \label{eqn:P2}
 P(\varsigma,d)= \frac{s_{\varsigma_D}}{\bar{x}_{\varsigma_D}\cdot \sqrt{|D|}}\times \frac{count(\varsigma_d)}{count(\varsigma_D)}
 \end{equation}
 \begin{equation}
 \label{eqn:R}
 R(x)=\frac{e^{x/10}-e^{-x/10}}{e^{x/10}+e^{-x/10}}
 \end{equation}
\begin{equation}
 \label{eqn:clust}\begin{split}
W_{edge}=\sum_i^{SET_d} \sum_{j\neq i}^{SET_d} \{Bool_{EM_n}(\varsigma_{\mathbf{e}i},\varsigma_{\mathbf{e}j}) \cdot \\ [\sum_{\varsigma_i=\varsigma_j}Rel(\varsigma,d)+cos(\mathbf{e}_i,\mathbf{e}_j)]\}
\end{split}
 \end{equation}
The penalty score ($P$) utilizes the coefficient of variation~\citep{Brown1998} to grasp the relative dispersion of stakeholder counts with respect to the sample mean across all topics. This value will be max-min normalized, and multiplied by the percentage of occurrence in the current topic. The boundary of $P$ and related proof will be in Appendix~\ref{appn:P-roof}. On the other hand, the reward score ($R$) uses a dampened hyperbolic tangent score with a factor of $\frac{1}{10}$ to limit the reward value from exploding, where the maximum possible value is capped at $1$. The final relevance score thus multiplies the reward with the penalty score, before using a hyperparameter $\beta$ to tune the significance of the relevance score.
 
To test the efficacy of the aforementioned penalty score, we introduce another penalty scoring, which is an altered version of the inverse document frequency (IDF) found in BM25~\citep{Robertson_Zaragoza_2009}. Unlike Equation~\ref{eqn:P2}, Equation~\ref{eqn:P1} directly uses the absence and presence of stakeholder counts across all documents to estimate the degree to penalize; this value however, fails to capture the actual extent of topic-based counts since it does not consider topic-based absence but document-based absence across the full document set. This may lead to edge cases being misrepresented: cases with common rarity and low reoccurrence (\textit{Irrelevant}) may have same penalty output as those with both rare and high reoccurrence (\textit{Significant}) since both cases provides high $|\forall A_\varsigma\in D|$ values. Contrary to using $P_{IDF}$, the coefficient of variation in the original penalty score represents the spread of $\varsigma$ across all topics, which allows the subsequent penalization of compact $\varsigma$ counts while identifying low and high frequency topics in high-variation $\varsigma$ counts through percentage multiplications. The behaviour of the different penalty and reward scores will be further illustrated in Appendix~\ref{appn:scores}.
\begin{equation}
 \label{eqn:P1}
 P_{IDF}(\varsigma)= lg(\frac{|\forall A\in D| - |\forall A_\varsigma\in D|+0.5}{|\forall A_\varsigma\in D|+0.5})
 \end{equation}

To increase the strictness of relevant events within the same cluster, we have also experimented with incorporating stakeholder Exact Matching (EM), where every 2 nodes requires at least N-matching unique stakeholders to have an edge. Therefore, the clustering process uses the encoded event summary from a General Text Embedding (GTE) Model~\citep{zhang2024mgtegeneralizedlongcontexttext} to obtain query-based cosine similarity scores combined with relevancy scores to generate the top 20 similar events for every node. These events have to satisfy the boolean EM condition to gain a weighted connection ($W_{edge}$) with the query node (Eqn.~\ref{eqn:clust}). Due to the use of relevancy scores, our method does not need to incorporate resource-intensive pairwise LLM comparisons of events found in the same cluster~\citep{hu-etal-2024-moments,Qorib_Hu_Ng_2025}.
%or self-questioning methods~\citep{wu-etal-2025-unfolding} 

\subsection{Timeline Generation}
After obtaining multiple clusters, each cluster $C$ undergoes a re-ranking based on both their size and the existing relevance of their events' stakeholders (Eqn.~\ref{eqn:tlg}); instead of incorporating relevance scoring to aid in edge connection (Eqn.~\ref{eqn:clust}), the relevance score is applied to determine how significant each cluster is relative to both $|C|$ and $Rel$. Events which are measured as significant will then be passed into TextRank~\citep{textrank} to identify important nodes within existing clusters. The final set of nodes will subsequently be used in the final Timeline Generation (TLG).
 \begin{equation}
 \label{eqn:sc}
\mathcal{S}_C = \bigcup_{\mathbf{e} \in C} \left\{ \varsigma \mid \varsigma \in \mathcal{S}_\mathbf{e} \right\}
 \end{equation}

 \begin{equation}
 \label{eqn:tlg}
 Significance= [1+ln(|C|)]\cdot \frac{\sum_{\varsigma\in\mathcal{S}_C} Rel(\varsigma,d)}{|\mathcal{S}_C|}
 \end{equation}

Henceforth, variations of overlapping events over different articles will not be a pressing concern since the clustering process is able to identify these variations based both the $cos$ score from similar texts and their $Rel+EM$ scores from repeated stakeholders to create a tighter cluster. In addition, TextRank, which causes nodes which are more different but within the same cluster to be used, would balance the variations in duplicates within their own cluster to better represent the events which are same with differing narrations. The penalty represented in $Rel$ also helps prevent similar but different events from falling into wrong cluster.

\section{Experimental Setups} \label{sec:general-setups}

\paragraph{Datasets.} 
We used two renowned TLS datasets, Timeline17 (T17)~\citep{t17} and Crisis~\citep{crisis}. T17 contains 19 timelines compiled from varying sources of online news sites, spanning 9 major topics from 2005-2013, each with 1-5 ground truth timelines. On the other hand,  Crisis has 22 annotated timelines covering 4 critical crisis events, each containing 4-7 ground truth timelines.

\paragraph{Benchmark.} We use two recent SOTA papers as our baselines: \citeauthor{hu-etal-2024-moments} and \citeauthor{wu-etal-2025-unfolding} are the current best-performing works and have reproducible results on T17 and Crisis, beating prior popular methods such as CLUST~\citep{gholipour-ghalandari-ifrim-2020-examining} and EGC~\citep{li-etal-2021-timeline}.

\paragraph{Models and Deployment.} For experiments, we adopted Qwen2.5-72B-Instruct~\citep{qwen2025qwen25technicalreport}, and GPT-4 omni~\citep{openai2024gpt4ocard}. We used GTE-Modernbert-Base for encoding. For the main experimental results, we used the best model, Qwen2.5-72B-Instruct to compare across the different benchmarks. All model deployment were done with VLLM~\citep{kwon2023efficient} with 8 H200 GPUs. We used batch sizes of 256 for batch encoding, and limit each event summary to a maximum of 4096 characters.

\paragraph{Metrics.} 
Adhering to previous work done~\citep{hu-etal-2024-moments,Qorib_Hu_Ng_2025,wu-etal-2025-unfolding}, we incorporate a part of the Tilse framework~\citep{steen-markert-2019-abstractive} and use three main scores to analyse the performance of the TLS task. We use an Alignment-based ROUGE-1 F1-score (AR-1) to evaluate the semantic distance of unigram overlaps between generated timelines and the provided ground truth. We also used the corresponding bigram overlaps and scored the Alignment-based ROUGE-2 F1-score (AR-2). Lastly, the Date-F1 metric will score the similarity of the dates in the generated timeline compared to the referenced ground truth to understand the quality of major events picked up.

\paragraph{Setups.} Following prior papers~\citep{hu-etal-2024-moments,Qorib_Hu_Ng_2025}, the TLS task can be defined as follows: given a set of temporally labelled news articles that is related to a broad topic ($d\rightarrow{A_1,A_2,A_3...}, \forall d\in D$), as well as the expected number of dates and the number of sentences to include in each date, a single timeline summarizing the topic should be generated. This generated timeline will subsequently be compared with the ground truth provided in the datasets. The baselines from the selected benchmarks are run in their publicly available Github repository. We have also published~\href{https://github.com/Veracitea/SUnSET}{\textcolor{purple}{\textit{SUnSET}}} publicly on Github for reproduction.

%https://github.com/Veracitea/SUnSET
%https://github.com/for-anonymous-reviewer/SUnSET/tree/main

\section{Experimental Results}
Table~\ref{tab:performance} showcases SUnSET as compared to the previously SOTA baselines. We selected the strongest model for each method for the comparisons. We also included Llama2-13B in Table~\ref{tab:performance} since Llama2 was the original model used in LLM-TLS's paper~\citep{hu-etal-2024-moments} and it occasionally performs better than Qwen2.5-72B. %%%%%%%%%%%%%%%%%%%%%%%%%%%%%%%%%%%%%%%%%%%%%%%%%%%%%%%%
\begin{table}[ht]
\centering
{
\setlength{\tabcolsep}{4pt}
\resizebox{0.49\textwidth}{!}{
\begin{tabular}{llcccc}
\toprule
\textbf{Dataset} & \textbf{Method} & \textbf{LLM} & \textbf{AR-1} & \textbf{AR-2} & \textbf{Date-F1} \\
\midrule

\multirow{6}{*}{\textbf{Crisis}} 
  & CHRONOS     & Qwen72B      & 0.108 & 0.045 & 0.323 \\
  & LLM-TLS & Llama13B & 0.112 & 0.032 & 0.329 \\
  & LLM-TLS & Qwen72B     & 0.111 & 0.036 & 0.326 \\
  \cmidrule(lr){2-6}
  & SUnSET  & Qwen72B & \textbf{0.129} & \textbf{0.047} & \textbf{0.389}\\ 
    & SUnSET  & GPT-4o & 0.107 & 0.036 & 0.381 \\

\midrule

\multirow{6}{*}{\textbf{T17}}
  & CHRONOS & Qwen72B  & 0.116 & 0.042 & 0.522 \\
  & LLM-TLS & Llama13B & 0.118 & 0.036 & 0.528 \\
  & LLM-TLS & Qwen72B & 0.114 & 0.040 & 0.543 \\
  \cmidrule(lr){2-6}
  & SUnSET  & Qwen72B & \textbf{0.136} & \textbf{0.044} &  0.576 \\
  &SUnSET  & GPT-4o & 0.120 & 0.039 & \textbf{0.590} \\

\bottomrule
\end{tabular}}
\caption{Performance of SUnSET on Crisis and T17 datasets}

\label{tab:performance}
}
\end{table}
%%%%%%%%%%%%%%%%%%%%%%%%%%%%%%%%%%%%%%%%%%%%%%%%%%%%%%%%%

Our method is able to defeat all prior work in all our metrics used for evaluation. For instance, SUnSET managed to improve Crisis Date-F1 score by $18.2\%$, ROUGE-1 by $15.2\%$ and ROUGE-2 by $30.6\%$ as compared to the previous SOTA result. The increase in Date-F1 scores over all prior methods indicates the superiority of SUnSET in correctly identifying important dates that were curated in the ground truth, whereas the high ROUGE scores support SUnSET in its ability to retain important details in the distillation process.  SUnSET also managed to obtain the best scores in T17, surpassing the previous SOTA. The full set of experimental results are included in Appendix~\ref{appn:expf}; examples of timeline output can be accessed in Appendix~\ref{appn:example}. For reference, we attempted using SUnSET with a closed-sourced model (GPT-4omni). GPT-4o performed relatively well and was able to have strong improvements in Date-F1 scores, but lacks in its ROUGE scores as compared to Qwen experiments. To understand where the difference in improvement comes from, more experiments were done on the clustering and timeline generation processes. 

\subsection{Relevancy in Event Clustering}
\begin{figure*}[ht]
  \centering
  \includegraphics[width=1.01\linewidth]{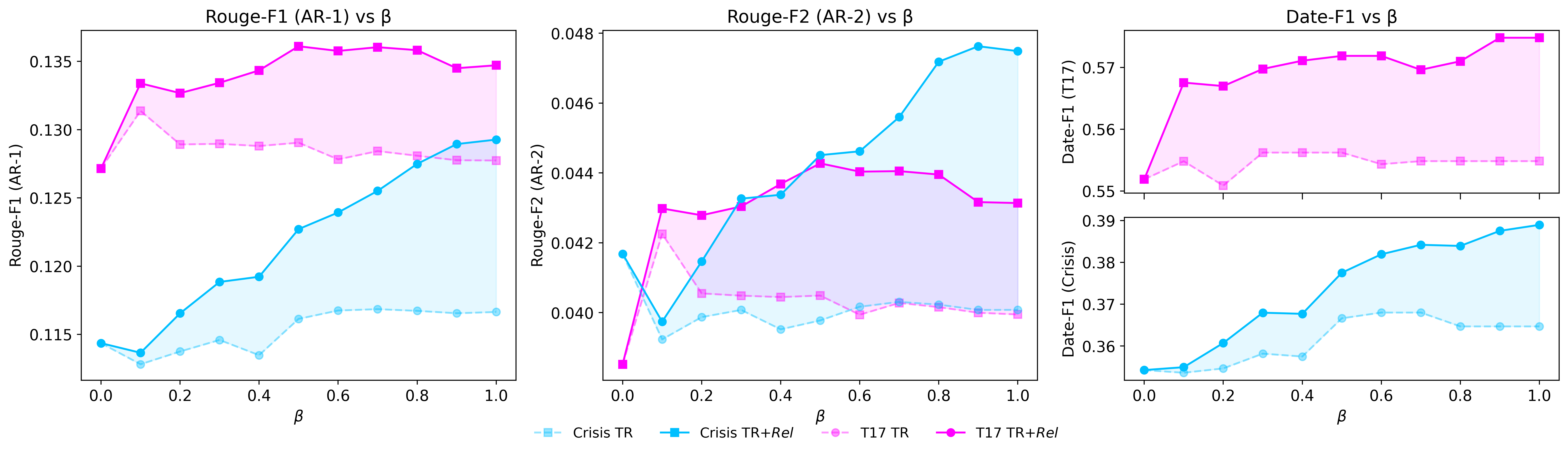}
  \caption{Effect of $\beta$ hyperparameter on $Rel$ in event clustering (TR) and $Rel$ in both event clustering and timeline generation (TR+$Rel$). The leftmost graph compares ROUGE-F1 values, the middle graph compares ROUGE-F2 values, and the rightmost graphs compare Date-F1 values.}
  \label{fig:beta}
\end{figure*}

\begin{table}[ht]
\centering
\caption{$Rel$ in Event Clustering}
\resizebox{0.4\textwidth}{!}{
\begin{tabular}{lccccc}
\toprule
\textbf{Dataset} & \textbf{Method} & \textbf{AR-1} & \textbf{AR-2} & \textbf{Date-F1} \\
\midrule
\multirow{2}{*}{\textbf{Crisis}} 
 & w/o $Rel$ & 0.114 & 0.041 & 0.354 \\
 & $Rel$ & \textbf{0.129} & \textbf{0.047} & \textbf{0.389} \\
\midrule
\multirow{2}{*}{\textbf{T17}}
 & w/o $Rel$ & 0.127 & 0.038 & 0.551 \\
 & $Rel$ & \textbf{0.136} & \textbf{0.044} & \textbf{0.576} \\

\bottomrule
\end{tabular}}
\label{tab:relclus}
\end{table}

We further investigated the impact of relevance used in the event clustering process by ablating all stakeholder-related information and relevancy scores, retaining only the cosine similarity scores of the events. From the results observed in Table~\ref{tab:relclus}, the use of relevance scoring and stakeholder information causes a leap in performance on Date-F1 scores by $10\%$ for Crisis and $4.5\%$ for T17. This indicates that our method for the identification of stakeholders plays an integral part in ranking event importance. 

Moreover, even without implementing $Rel$, by comparing Table~\ref{tab:performance}'s LLM-TLS score and Table~\ref{tab:relclus}'s ``w/o $Rel$'', it illustrates the significance of the event component (E). It is observed that utilizing LLMs to generate event sets $\mathbf{e}$ in SET is already more effective than existing methods which typically uses summarization of $A$~\citep{Qorib_Hu_Ng_2025,hu-etal-2024-moments} or a self-questioning rewriter~\citep{wu-etal-2025-unfolding}. Even without pairwise matching (``w/o $Rel$''), using only the cosine similarity of $\mathbf{e}$ managed to beat the benchmark set by LLM-TLS and CHRONOS (Tab.~\ref{tab:performance}). This reveals that extraction of multiple mentioned events within a single article is vital for TLS tasks; this is intuitive, as news articles often reference related developments that connect to the central event being reported. Significant occurrences may receive limited coverage initially, until a subsequent development---part of an ongoing sequence---draws renewed attention from multiple news outlets. Therefore, since the time component in SET is always a necessary input due to the nature of this task, \textit{we have shown that all components in SET are important for the performance of timeline generation.}

\subsection{Relevancy in Timeline Generation}

As stakeholder relevance was utilized in the Timeline Generation process, we compared the difference between the use and absence of $Rel$ to estimate the significance of each cluster before passing them into TextRank . We used the best performing results for both settings (Tab.~\ref{tab:reltlg}). The inclusion of relevance scoring improved the performance across all metrics for both Crisis and T17; this demonstrates the efficacy of utilizing $Rel$ for cluster ranking, which ultimately improved the selection of important dates while retaining the details of important events. 

\begin{table}[ht]
\centering
\caption{$Rel$ in Timeline Generation}
\resizebox{0.45\textwidth}{!}{
\begin{tabular}{lccccc}
\toprule
\textbf{Dataset} & \textbf{Method} & \textbf{AR-1} & \textbf{AR-2} & \textbf{Date-F1} \\
\midrule
\multirow{2}{*}{\textbf{Crisis}} 
 & TextRank & 0.117 & 0.040 & 0.368 \\
 & TextRank+$Rel$ & \textbf{0.129} & \textbf{0.047} & \textbf{0.389} \\
\midrule
\multirow{2}{*}{\textbf{T17}}
 & TextRank & 0.128 & 0.041 & 0.559 \\
 & TextRank+$Rel$ & \textbf{0.136} & \textbf{0.044} & \textbf{0.576} \\

\bottomrule
\end{tabular}}
\label{tab:reltlg}
\end{table}

To go a step further, we looked into the influence of the $\beta$ hyperparameter over the entire SUnSET method, where we compared the inclusion and exclusion of $Rel$ during the timeline generation process (Fig.~\ref{fig:beta}). In the full SUnSET process (TextRank+$Rel$), all three scores (ROUGE-F1, ROUGE-F2 and Date-F1) increases as $\beta$ increases. This behaviour is more latent when $Rel$ is only used in clustering (TextRank only) as a subsequent decrease in performance is observed as $\beta$ increases beyond plateau. 

There are several differences in Crisis and T17's increase as $\beta$ grows. T17, which contains a larger amount of topics but a smaller pool of articles within each topic (larger $D$ and smaller $A$) tends to have a sharp and significant increase with a small $\beta$ at $0.1$; although it shows the use of $Rel$ is effective once introduced, the performance may subsequently become stagnant with less significant changes with the increment of $\beta$. In contrary, Crisis with well established topics containing large pools of articles tend to benefit more as $\beta$ increases after its initial introduction, where small $\beta$ values are not helpful and potentially worsens performance, and larger $\beta$ values provide remarkable growth.

\subsection{Penalty PvP: $P$ and $P_{IDF}$}

\begin{table}[ht]
\centering
\caption{Penalty PvP- $P$ versus $P_{IDF}$}
\resizebox{0.4\textwidth}{!}{
\begin{tabular}{lccccc}
\toprule
\textbf{Dataset} & \textbf{Penalty} & \textbf{AR-1} & \textbf{AR-2} & \textbf{Date-F1} \\
\midrule
\multirow{2}{*}{\textbf{Crisis}} 
 & $P_{IDF}$ & 0.119 & 0.043 & 0.374 \\
 & $P$ & \textbf{0.129} & \textbf{0.047} & \textbf{0.389} \\
\midrule
\multirow{2}{*}{\textbf{T17}}
 & $P_{IDF}$ & \textbf{0.137} & 0.043 & \textbf{0.576} \\
 & $P$ & 0.136 & \textbf{0.044} & \textbf{0.576} \\

\bottomrule
\end{tabular}}
\label{tab:PvP}
\end{table}

As we introduced a total of two different penalty scores, we tested the difference in the type of penalty used within the relevance score for both datasets. Table~\ref{tab:PvP} shows the difference in using $P$ and $P_{IDF}$ for SUnSET, where the best result out of all $\beta$ attempted was used. We observed that the impact on T17, which contains smaller datasets and thus lesser stakeholders than Crisis, is smaller; the use of either $P$ and $P_{IDF}$ causes little differences in T17's final results, yet it causes significant differences in a larger dataset such as Crisis. This indicates that $P$ is more flexible in more cases, especially when there exist a larger event space where more \textit{Significant} and \textit{Irrelevant} stakeholders (Tab.~\ref{tab:freq_table}) can be identified.

\subsection{Stakeholder Exact Matching}

\begin{table}[ht]
\centering
\caption{Performance of using EM in Event Clustering; best $\beta$ values are recorded when $\beta>0$}
\resizebox{0.4\textwidth}{!}{
\begin{tabular}{lcccccc}
\toprule
\textbf{Dataset} & \textbf{$\beta$} & \textbf{EM} & \textbf{AR-1} & \textbf{AR-2} & \textbf{Date-F1} \\
\midrule
\multirow{6}{*}{\textbf{Crisis}} 
 & \multirow{3}{*}{$0$} & $0$ &  0.114 & 0.041 & 0.354  \\
 && $1$ & \textbf{0.119} & \textbf{0.044} & \textbf{0.364} \\
 && $2$ & 0.114 & 0.041 & 0.358 \\
\cmidrule(lr){2-6}
 & 1.0 & $0$ & \textbf{0.129} & \textbf{0.047} & \textbf{0.389} \\
 & 0.9 & $1$ & 0.127 & 0.046 & 0.386 \\
 & 0.6 & $2$ & 0.121 & 0.042 & 0.379 \\
\midrule
\multirow{3}{*}{\textbf{T17}}
 & \multirow{3}{*}{$0$} & $0$ & 0.127 & 0.038 & 0.553 \\
 && $1$ & \textbf{0.128} & \textbf{0.040} & 0.553 \\
 && $2$ & 0.127 & \textbf{0.040} & \textbf{0.558} \\
\cmidrule(lr){2-6}
 & 1.0 & $0$ & 0.134 & 0.043 & 0.574 \\
 & 0.9 & $1$ & \textbf{0.136} & \textbf{0.044} & \textbf{0.576} \\
 & 1.0 & $2$ & 0.134 & 0.044 & 0.573 \\

\bottomrule
\end{tabular}}
\label{tab:EM}
\end{table}

Our initial experiments used only EM with SET without consideration of relevance scores. In Table~\ref{tab:EM}, when $\beta$ is set as 0, there is no relevance scoring used, and we observe that the inclusion of EM is able to capture meaningful stakeholder information to a proportional degree. Typically, $EM_1$ is sufficient to result in an improvement, and further matching counts may instead deteriorate the performance of models. This may be due to insufficient stakeholders with high counts, where the median of repetitive stakeholders typically lies around 3 across all datasets (Appn.~\ref{appn:stakeC}). Though EM performance does not match up to the addition of $Rel$, this is sufficient to signify the importance of stakeholder information in the TLS. 

Furthermore, the improvement of $Rel$ over EM illustrates the strength of our formulated equation in stakeholder ranking. While $Rel$ is used, the use of EM causes limited improvements as $Rel$ in itself is able to capture and rank the entities involved in $\mathbf{e}$. It is also observed that high $\beta$ values typically performs better across differing EM settings, which further emphasizes the importance of $Rel$.

\subsection{Time Comparison}

\begin{table}[ht]
\centering
\caption{GPU hours for clustering}
\resizebox{0.35\textwidth}{!}{
\begin{tabular}{lcc}
\toprule
\textbf{Dataset} & \textbf{Method} & \textbf{GPU Time} \\
\midrule
\multirow{3}{*}{\textbf{Crisis}} 
 & LLM-TLS & 7 hours 12 min \\
  & CHRONOS & 24 min \\
    \cmidrule(lr){2-3}
 & SUnSET & 7 \textbf{\textit{min}} 14 sec \\
\midrule
\multirow{2}{*}{\textbf{T17}}
 & LLM-TLS & 2 hours 12 min \\
  & CHRONOS & 1 hour 9 min \\
    \cmidrule(lr){2-3}
 & SUnSET & 1 \textbf{\textit{min}} 41 sec \\
 
\bottomrule
\end{tabular}}
\end{table}

Lastly, we measured the time required for running the experiments, and realised the high efficiency for using SUnSET as compared to previous SOTA methods. The benefits mainly comes from the lack of additional LLM prompting beyond SET creation, where CHRONOS utilizes questioning rounds and LLM-TLS prompts once for every pairwise comparison required. Therefore, SUnSET is able to bypass such time-consuming prompt-questioning methods due to the high efficacy of utilizing stakeholder information for distilling events into a compact representation. 

\section{Conclusions}
We propose SUnSET, a novel approach that leverages SET triplets and stakeholder relevance to generate milestone events for TLS problems. By incorporating stakeholder-relevancy heuristics, SUnSET effectively addresses the challenges of distilling multiple event sets from complex articles. Our framework systematically tackles these issues in three critical steps: 1) SET generation expands the pool of candidate events per article; 2) SUnSET's efficient event clustering eliminates the need for time-intensive reclustering, even with larger candidate sets; 3) Stakeholder information encoding via $Rel$ enhances both clustering and timeline generation by yielding a more representative set of linked events. Through extensive experiments and ablation studies on the T17 and Crisis datasets, SUnSET consistently delivers state-of-the-art results, validating its effectiveness in both performance and interpretability. These findings underscore the potential of stakeholder-aware TLS systems to advance narrative coherence, relevance, and human-aligned summarization across evolving news domains, while also establishing stakeholder scoring as a scalable framework for prioritizing information in other contexts where audience-centric relevance and societal impact are key --- such as search, recommendation, and policy communication.

\section*{Limitations}
Our study advances TLS through the integration of SET and stakeholder ranking mechanisms, though certain limitations remain. Within our domain of control, we applied feasible refinement techniques such as prompt engineering and performed rigorous reviews to minimize inconsistencies. However, as with prior work, we were unable to fully mitigate hallucinations or factual inaccuracies that may arise during event extraction when utilizing LLMs due to their inherent generation behavior. Finally, in line with previous studies, we evaluated SUnSET using publicly available research datasets, without extending analysis to real-time news data with more dynamic, evolving topics. We identify this as a promising direction for future exploration.

\section*{Acknowledgments}
We would like to thank Austen Jeremy Sugiarto and Leong Chee Kai in aiding the initial stages of the experiments. We are also grateful to Dr. Basura Fernando for his help in reading through the paper. We deeply appreciate any reviewers for their helpful feedback. 

% Bibliography entries for the entire Anthology, followed by custom entries
%\bibliography{anthology,custom}
% Custom bibliography entries only
\bibliography{custom}

@misc{Statista, title={News and magazines - worldwide: Statista market forecast}, url={https://www.statista.com/outlook/amo/app/news-magazines/worldwide}, journal={Statista}, author={Statista}}

@Inbook{Brown1998,
author="Brown, Charles E.",
title="Coefficient of Variation",
bookTitle="Applied Multivariate Statistics in Geohydrology and Related Sciences",
year="1998",
publisher="Springer Berlin Heidelberg",
address="Berlin, Heidelberg",
pages="155--157",
isbn="978-3-642-80328-4",
doi="10.1007/978-3-642-80328-4_13",
url="https://doi.org/10.1007/978-3-642-80328-4_13"
}

@misc{zhang2024mgtegeneralizedlongcontexttext,
      title={mGTE: Generalized Long-Context Text Representation and Reranking Models for Multilingual Text Retrieval}, 
      author={Xin Zhang and Yanzhao Zhang and Dingkun Long and Wen Xie and Ziqi Dai and Jialong Tang and Huan Lin and Baosong Yang and Pengjun Xie and Fei Huang and Meishan Zhang and Wenjie Li and Min Zhang},
      year={2024},
      eprint={2407.19669},
      archivePrefix={arXiv},
      primaryClass={cs.CL},
      url={https://arxiv.org/abs/2407.19669}, 
}

@inproceedings{ribeiro-etal-2017-unsupervised,
    title = "Unsupervised Event Clustering and Aggregation from Newswire and Web Articles",
    author = "Ribeiro, Swen  and
      Ferret, Olivier  and
      Tannier, Xavier",
    editor = "Popescu, Octavian  and
      Strapparava, Carlo",
    booktitle = "Proceedings of the 2017 {EMNLP} Workshop: Natural Language Processing meets Journalism",
    month = sep,
    year = "2017",
    address = "Copenhagen, Denmark",
    publisher = "Association for Computational Linguistics",
    url = "https://aclanthology.org/W17-4211/",
    doi = "10.18653/v1/W17-4211",
    pages = "62--67"
}

@article{Robertson_Zaragoza_2009, title={The probabilistic relevance framework: BM25 and beyond}, volume={3}, DOI={10.1561/1500000019}, number={4}, journal={Foundations and Trends in Information Retrieval}, author={Robertson, Stephen and Zaragoza, Hugo}, year={2009}, pages={333–389}}

@article{Arnold_Goldschmitt_Rigotti_2023, title={Dealing with information overload: A comprehensive review}, volume={14}, DOI={10.3389/fpsyg.2023.1122200}, journal={Frontiers in Psychology}, author={Arnold, Miriam and Goldschmitt, Mascha and Rigotti, Thomas}, year={2023}, month={Jun}}

@misc{piryani2025itshightimesurvey,
      title={It's High Time: A Survey of Temporal Information Retrieval and Question Answering}, 
      author={Bhawna Piryani and Abdelrahman Abdallah and Jamshid Mozafari and Avishek Anand and Adam Jatowt},
      year={2025},
      eprint={2505.20243},
      archivePrefix={arXiv},
      primaryClass={cs.CL},
      url={https://arxiv.org/abs/2505.20243}, 
}

@misc{xiong2024largelanguagemodelslearn,
      title={Large Language Models Can Learn Temporal Reasoning}, 
      author={Siheng Xiong and Ali Payani and Ramana Kompella and Faramarz Fekri},
      year={2024},
      eprint={2401.06853},
      archivePrefix={arXiv},
      primaryClass={cs.CL},
      url={https://arxiv.org/abs/2401.06853}, 
}

@misc{qwen2025qwen25technicalreport,
      title={Qwen2.5 Technical Report}, 
      author={Qwen and An Yang and Baosong Yang and Beichen Zhang and Binyuan Hui and Bo Zheng and Bowen Yu and Chengyuan Li and Dayiheng Liu and Fei Huang and Haoran Wei and Huan Lin and Jian Yang and Jianhong Tu and Jianwei Zhang and Jianxin Yang and Jiaxi Yang and Jingren Zhou and Junyang Lin and Kai Dang and Keming Lu and Keqin Bao and Kexin Yang and Le Yu and Mei Li and Mingfeng Xue and Pei Zhang and Qin Zhu and Rui Men and Runji Lin and Tianhao Li and Tianyi Tang and Tingyu Xia and Xingzhang Ren and Xuancheng Ren and Yang Fan and Yang Su and Yichang Zhang and Yu Wan and Yuqiong Liu and Zeyu Cui and Zhenru Zhang and Zihan Qiu},
      year={2025},
      eprint={2412.15115},
      archivePrefix={arXiv},
      primaryClass={cs.CL},
      url={https://arxiv.org/abs/2412.15115}, 
}

@article{wikidata, title={Wikidata}, volume={57}, DOI={10.1145/2629489}, number={10}, journal={Communications of the ACM}, author={Vrandečić, Denny and Krötzsch, Markus}, year={2014}, month={Sep}, pages={78–85}}

@inproceedings{steen-markert-2019-abstractive,
    title = "Abstractive Timeline Summarization",
    author = "Steen, Julius  and
      Markert, Katja",
    editor = "Wang, Lu  and
      Cheung, Jackie Chi Kit  and
      Carenini, Giuseppe  and
      Liu, Fei",
    booktitle = "Proceedings of the 2nd Workshop on New Frontiers in Summarization",
    month = nov,
    year = "2019",
    address = "Hong Kong, China",
    publisher = "Association for Computational Linguistics",
    url = "https://aclanthology.org/D19-5403/",
    doi = "10.18653/v1/D19-5403",
    pages = "21--31"
}

@InProceedings{t17,
title = "Leveraging Learning To Rank in an Optimization Framework for Timeline Summarization",
author = "Tran, {Giang Binh} and Tuan Tran and Nam-Khanh Tran and Mohammad Alrifai and Nattiya Kanhabua",
year = "2013",
language = "Udefineret/Ukendt",
booktitle = "SIGIR 2013 Workshop on Time-aware Information Access (TAIA'2013)",
}

@InProceedings{crisis,
author="Tran, {Giang Binh}
and Alrifai, Mohammad
and Herder, Eelco",
editor="Hanbury, Allan
and Kazai, Gabriella
and Rauber, Andreas
and Fuhr, Norbert",
title="Timeline Summarization from Relevant Headlines",
booktitle="Advances in Information Retrieval",
year="2015",
publisher="Springer International Publishing",
address="Cham",
pages="245--256",
isbn="978-3-319-16354-3"
}

@inproceedings{gholipour-ghalandari-ifrim-2020-examining,
    title = "Examining the State-of-the-Art in News Timeline Summarization",
    author = "Gholipour Ghalandari, Demian  and
      Ifrim, Georgiana",
    editor = "Jurafsky, Dan  and
      Chai, Joyce  and
      Schluter, Natalie  and
      Tetreault, Joel",
    booktitle = "Proceedings of the 58th Annual Meeting of the Association for Computational Linguistics",
    month = jul,
    year = "2020",
    address = "Online",
    publisher = "Association for Computational Linguistics",
    url = "https://aclanthology.org/2020.acl-main.122/",
    doi = "10.18653/v1/2020.acl-main.122",
    pages = "1322--1334"
}

@inproceedings{hu-etal-2024-moments,
    title = "From Moments to Milestones: Incremental Timeline Summarization Leveraging Large Language Models",
    author = "Hu, Qisheng  and
      Moon, Geonsik  and
      Ng, Hwee Tou",
    editor = "Ku, Lun-Wei  and
      Martins, Andre  and
      Srikumar, Vivek",
    booktitle = "Proceedings of the 62nd Annual Meeting of the Association for Computational Linguistics (Volume 1: Long Papers)",
    month = aug,
    year = "2024",
    address = "Bangkok, Thailand",
    publisher = "Association for Computational Linguistics",
    url = "https://aclanthology.org/2024.acl-long.390/",
    doi = "10.18653/v1/2024.acl-long.390",
    pages = "7232--7246"
}

@article{Qorib_Hu_Ng_2025, title={Just What You Desire: Constrained Timeline Summarization with Self-Reflection for Enhanced Relevance}, volume={39}, url={https://ojs.aaai.org/index.php/AAAI/article/view/34691}, DOI={10.1609/aaai.v39i23.34691}, abstractNote={}, number={23}, journal={Proceedings of the AAAI Conference on Artificial Intelligence}, author={Qorib, Muhammad Reza and Hu, Qisheng and Ng, Hwee Tou}, year={2025}, month={Apr.}, pages={25065-25073} }

@inproceedings{wu-etal-2025-unfolding,
    title = "Unfolding the Headline: Iterative Self-Questioning for News Retrieval and Timeline Summarization",
    author = "Wu, Weiqi  and
      Huang, Shen  and
      Jiang, Yong  and
      Xie, Pengjun  and
      Huang, Fei  and
      Zhao, Hai",
    editor = "Chiruzzo, Luis  and
      Ritter, Alan  and
      Wang, Lu",
    booktitle = "Findings of the Association for Computational Linguistics: NAACL 2025",
    month = apr,
    year = "2025",
    address = "Albuquerque, New Mexico",
    publisher = "Association for Computational Linguistics",
    url = "https://aclanthology.org/2025.findings-naacl.248/",
    doi = "10.18653/v1/2025.findings-naacl.248",
    pages = "4385--4398",
    ISBN = "979-8-89176-195-7"
}

@inproceedings{li-etal-2021-timeline,
    title = "Timeline Summarization based on Event Graph Compression via Time-Aware Optimal Transport",
    author = "Li, Manling  and
      Ma, Tengfei  and
      Yu, Mo  and
      Wu, Lingfei  and
      Gao, Tian  and
      Ji, Heng  and
      McKeown, Kathleen",
    editor = "Moens, Marie-Francine  and
      Huang, Xuanjing  and
      Specia, Lucia  and
      Yih, Scott Wen-tau",
    booktitle = "Proceedings of the 2021 Conference on Empirical Methods in Natural Language Processing",
    month = nov,
    year = "2021",
    address = "Online and Punta Cana, Dominican Republic",
    publisher = "Association for Computational Linguistics",
    url = "https://aclanthology.org/2021.emnlp-main.519/",
    doi = "10.18653/v1/2021.emnlp-main.519",
    pages = "6443--6456"
}

@misc{chen2023followtimelinegeneratingabstractive,
      title={Follow the Timeline! Generating Abstractive and Extractive Timeline Summary in Chronological Order}, 
      author={Xiuying Chen and Mingzhe Li and Shen Gao and Zhangming Chan and Dongyan Zhao and Xin Gao and Xiangliang Zhang and Rui Yan},
      year={2023},
      eprint={2301.00867},
      archivePrefix={arXiv},
      primaryClass={cs.CL},
      url={https://arxiv.org/abs/2301.00867}, 
}

@inproceedings{textrank,
    title = "{T}ext{R}ank: Bringing Order into Text",
    author = "Mihalcea, Rada  and
      Tarau, Paul",
    editor = "Lin, Dekang  and
      Wu, Dekai",
    booktitle = "Proceedings of the 2004 Conference on Empirical Methods in Natural Language Processing",
    month = jul,
    year = "2004",
    address = "Barcelona, Spain",
    publisher = "Association for Computational Linguistics",
    url = "https://aclanthology.org/W04-3252/",
    pages = "404--411"
}

@inproceedings{2004paper,
author = {Chieu, Hai Leong and Lee, Yoong Keok},
title = {Query based event extraction along a timeline},
year = {2004},
isbn = {1581138814},
publisher = {Association for Computing Machinery},
address = {New York, NY, USA},
url = {https://doi.org/10.1145/1008992.1009065},
doi = {10.1145/1008992.1009065},
booktitle = {Proceedings of the 27th Annual International ACM SIGIR Conference on Research and Development in Information Retrieval},
pages = {425–432},
numpages = {8},
location = {Sheffield, United Kingdom},
series = {SIGIR '04}
}

@misc{wang2023webnewstimelinegeneration,
      title={Web News Timeline Generation with Extended Task Prompting}, 
      author={Sha Wang and Yuchen Li and Hanhua Xiao and Lambert Deng and Yanfei Dong},
      year={2023},
      eprint={2311.11652},
      archivePrefix={arXiv},
      primaryClass={cs.AI},
      url={https://arxiv.org/abs/2311.11652}, 
}

@inproceedings{sojitra,
author = {Sojitra, Daivik and Jain, Raghav and Saha, Sriparna and Jatowt, Adam and Gupta, Manish},
title = {Timeline Summarization in the Era of LLMs},
year = {2024},
isbn = {9798400704314},
publisher = {Association for Computing Machinery},
address = {New York, NY, USA},
url = {https://doi.org/10.1145/3626772.3657899},
doi = {10.1145/3626772.3657899},
booktitle = {Proceedings of the 47th International ACM SIGIR Conference on Research and Development in Information Retrieval},
pages = {2657–2661},
numpages = {5},
location = {Washington DC, USA},
series = {SIGIR '24}
}

@misc{openai2024gpt4ocard,
      title={GPT-4o System Card}, 
      author = {OpenAI},
      year={2024},
      eprint={2410.21276},
      archivePrefix={arXiv},
      primaryClass={cs.CL},
      url={https://arxiv.org/abs/2410.21276}, 
}

@inproceedings{kwon2023efficient,
  title={Efficient Memory Management for Large Language Model Serving with PagedAttention},
  author={Woosuk Kwon and Zhuohan Li and Siyuan Zhuang and Ying Sheng and Lianmin Zheng and Cody Hao Yu and Joseph E. Gonzalez and Hao Zhang and Ion Stoica},
  booktitle={Proceedings of the ACM SIGOPS 29th Symposium on Operating Systems Principles},
  year={2023}
}

@inproceedings{someart,
author = {Sojitra, Daivik and Jain, Raghav and Saha, Sriparna and Jatowt, Adam and Gupta, Manish},
title = {Timeline Summarization in the Era of LLMs},
year = {2024},
isbn = {9798400704314},
publisher = {Association for Computing Machinery},
address = {New York, NY, USA},
url = {https://doi.org/10.1145/3626772.3657899},
doi = {10.1145/3626772.3657899},
booktitle = {Proceedings of the 47th International ACM SIGIR Conference on Research and Development in Information Retrieval},
pages = {2657–2661},
numpages = {5},
location = {Washington DC, USA},
series = {SIGIR '24}
}

\newpage
\onecolumn 
\appendix
\section{Prompts used for LLMs}~\label{appn:prompts}
\begin{tcolorbox}[title=Prompt for Event and Time Generation, colback=gray!10, colframe=black!50, boxrule=0.5mm]
\begin{verbatim}
You are a professional journalist that is tasked to generate date-based
event summary of a given article. A single list contains an article and
its published time. You should generate a dictionary of the most
relevant events of an article, where each key in the dictionary is a 
string of the expected event start date in terms of Year-Month-Day 
(e.g.2011-12-25) and the value will be a summary of the relevant events 
on that day. Summarize only the most important events found in the 
article, as succinctly as possible. If you are uncertain of the date of 
an event, feel free to use the published date. You should only output
the dictionary in your answer. Generate a dictionary of events of the
following article: {str(article_x)}.
\end{verbatim}
\end{tcolorbox}

\begin{tcolorbox}[title=Prompt for Stakeholder Generation, colback=gray!10, colframe=black!50, boxrule=0.5mm]
\begin{verbatim}
You are a professional journalist that is tasked to generate the most 
relevant stakeholders relevant to a given event summary of an article. 
A single list contains an article and its published time. You should 
generate a singular list containing not more than five relevant 
stakeholders related to only the stipulated event mentioned. These 
stakeholders should not be general, and must be identifiable named 
entities that can be matched to a person, organization or role when read 
on its own. Every single stakeholder generated should also ideally exist 
in exact wording as mentioned within the original article. You should 
only output the list of stakeholders in your answer, and all 
stakeholders should be enclosed in string format. Generate a list of
related stakeholders of event: {dict[key_x]}.
Given article: {str(article_x)}.

\end{verbatim}
\end{tcolorbox}
\newpage

% \section{Building Knowledge Graph from Wikidata\footnote{Main API: \url{https://www.wikidata.org/wiki/Wikidata:REST_API}, NER module: \url{https://huggingface.co/spacy/en_core_web_trf}, Request module: \url{https://requests.readthedocs.io/en/latest/}}}~\label{appn:wikidata}

\section[Building Knowledge Graph from Wikidata]{Building Knowledge Graph from Wikidata\footnote{Main API: \url{https://www.wikidata.org/wiki/Wikidata:REST_API}, NER module: \url{https://huggingface.co/spacy/en_core_web_trf}, Request module: \url{https://requests.readthedocs.io/en/latest/}}} \label{appn:wikidata}

\begin{algorithm}
\caption{Knowledge Graph for Coreference Resolution}
\begin{algorithmic}[1]
\STATE Initialize an empty dictionary: $d \gets \{\}$
\FOR{Every stakeholder $\varsigma$ in $\mathcal{S}$}
    \IF{$\varsigma$ in $d$}
        \STATE \textbf{continue} 
    \ELSE
        \STATE Search $\varsigma$ in Wikidata label/alt-label with API
        \IF{output $O$ does not exist}
            \STATE $\varsigma_\gamma \gets$ Remove title in $\varsigma$ with NER
            \STATE Search $\varsigma_\gamma$ in Wikidata label/alt-label with API
            \IF{output $O$ does not exist}
                \STATE $\varsigma_\delta \gets$ Remove whitespace and replace with \texttt{\&\&}
                \STATE Search $\varsigma_\gamma$ in Wikidata label/alt-label with API
                \IF{output $O$ does not exist}
                    \STATE Use request API for Wikidata interface search
                    \IF{output $O$ does not exist}
                        \STATE $d[\varsigma] \gets \varsigma$
                        \STATE \textbf{continue}
                    \ENDIF
                \ENDIF
            \ENDIF
        \ENDIF
    \ENDIF
    \IF{Operator "Position Held By" exist in $O$}
        \STATE $P \gets$ ID under "Position Held By"
        \STATE $d[\varsigma] \gets P$
        \STATE \textbf{continue}
    \ELSE
        \STATE $d[\varsigma] \gets O$
        \STATE \textbf{continue}
    \ENDIF
\ENDFOR
\end{algorithmic}
\end{algorithm}

\newpage

\section{Boundary for Penalty Score}~\label{appn:P-roof}
\begin{proof}
Let $x \in \mathbb{Z}$ and $x \geq 0$ since $x$ represents counts. $|D|$ represents the size of all topics.
\begin{equation}
\label{eq:cv}
CV=\frac{s}{\bar{x}}
\end{equation}

From Equation~\eqref{eq:cv}, 
\begin{equation}
\label{eq:cv2}
CV^2=\frac{\frac{\sum(x_i-\bar{x})^2}{|D|-1}}{\bar{x}^2}=\frac{\frac{\sum x_i^2-2\bar{x}\sum x_i+\sum \bar{x}^2}{|D|-1}}{\bar{x}^2}=\frac{\frac{\sum x_i^2-2|D|\bar{x}^2+|D|\bar{x}^2}{|D|-1}}{\bar{x}^2}
\end{equation}

Since n>1 and $\bar{x}\geq0$,
\begin{equation}
\label{eq:lol}
\sum^{|D|}_{i=1}x^2_i\leq(\sum^{|D|}_{i=1}x_i)^2=|D|^2\bar{x}^2
\end{equation}

From Equation~\eqref{eq:lol}, minus $\bar{x}^2|D|$,
\begin{equation}
\label{eq:lol2}
\sum^{|D|}_{i=1}x^2_i-|D|\bar{x}^2\leq |D|^2\bar{x}^2-|D|\bar{x}^2
\end{equation}

Change Equation~\eqref{eq:lol2} LHS to Equation~\eqref{eq:cv2},
\begin{equation}
\label{eq:haha}
CV^2=\frac{\frac{\sum x_i^2-|D|\bar{x}^2}{|D|-1}}{\bar{x}^2}
\leq \frac{\frac{|D|^2\bar{x}^2-|D|\bar{x}^2}{|D|-1}}{\bar{x}^2}=|D|
\end{equation}

From Equation~\eqref{eq:haha}, since $n>1$ and $\bar{x}\geq0$,\\
\begin{center}
$0\leq CV \leq \sqrt{|D|}$
\end{center}

From Equation~\eqref{eqn:P2}, we divided $CV$ by $\sqrt{D}$, and multiplied it by the percentage of $\varsigma$ belonging in $d$, where $0\leq \frac{count(\varsigma_d)}{count(\varsigma_D)}\leq 1$, therefore,
\begin{center}
$0\leq P\leq 1$
\end{center}
\end{proof}

\newpage

\section{Behaviour of Reward and Penalty Scores}~\label{appn:scores}

\begin{figure*}[ht]
  \centering
  \includegraphics[width=\linewidth]{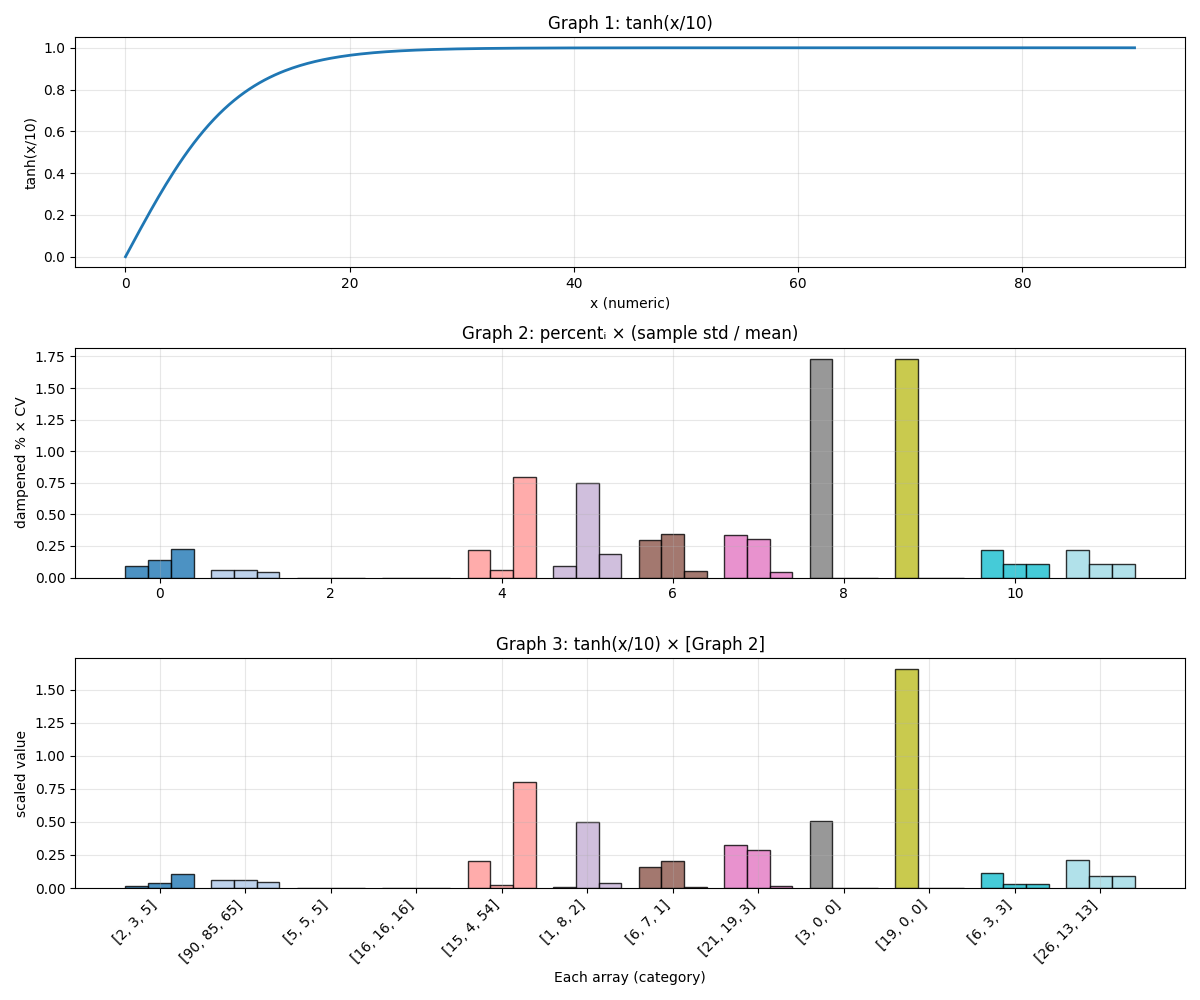}
  \caption{Case-by-case representation of Penalty Behaviour}
  \label{fig:pv2crisis}
\end{figure*}

We illustrate the behaviour of both Penalty and Reward in this section. From Figure~\ref{fig:pv2crisis}, Graph $1$ plots $R$ (Eqn.~\ref{eqn:R}) as the count of stakeholder increases. It can be observed that when the count of the stakeholders ($x\leftarrow count(\varsigma_d)$) reach approximately 21, the value of $R$ saturates regardless of any more addition to counts for the same stakeholder. Subsequently, Graph $2$ plots the equation of unnormalized $P$ (Eqn.~\ref{eqn:P2}) of $12$ different scenarios (Tab.~\ref{tab:crisispv2sit}).

% Please add the following required packages to your document preamble:
% \usepackage[table,xcdraw]{xcolor}
% Beamer presentation requires \usepackage{colortbl} instead of \usepackage[table,xcdraw]{xcolor}
\begin{table}[H]
\begin{tabular}{|l|l|l|}
\hline
\rowcolor[HTML]{9B9B9B} 
\textit{\textbf{Index}} & \textbf{Stakeholder Counts} & \textbf{Rationale} \\ \hline
0 & A: 2, B: 3, C:5 & Close Distribution, All Low Counts \\ \hline
1 & A: 90, B: 85, C:65 & Close Distribution, All High Counts \\ \hline
2 & A: 5, B: 5, C:5 & Uniform Distribution, All Low Counts \\ \hline
3 & A: 16, B: 16, C:16 & Uniform Distribution, All High Counts \\ \hline
4 & A: 15, B: 4, C:54 & Single-Peak Distribution, Peak High Count \\ \hline
5 & A: 1, B: 8, C:2 & Single-Peak Distribution, Peak Low Counts \\ \hline
6 & A: 6, B: 7, C:1 & Double-Peaks Distribution, Peaks Low Counts \\ \hline
7 & A: 21, B: 19, C:3 & Double-Peaks Distribution, Peaks High Counts \\ \hline
8 & A: 3, B: 0, C:0 & Single-Peak Distribution, Peak Low Count, Rest are 0 \\ \hline
9 & A: 19, B: 0, C:0 & Single-Peak Distribution, Peak High Count, Rest are 0 \\ \hline
10 & A: 6, B: 3, C:3 & Single-Peak Distribution, Peak Low Count, Rest are half \\ \hline
11 & A: 26, B: 13, C:13 & Single-Peak Distribution, Peak High Count, Rest are half \\ \hline
\end{tabular}
\caption{Scenarios used for Stakeholder Counts across Three Topics (A, B and C)}
\label{tab:crisispv2sit}
\end{table}

Graph $3$ plots the value of $Rel$ without consideration of $\beta$ (by setting $\beta=1$) to understand the impact of multiplying $P$ and $R$. Cases 0, 1, 2, 3, 10 and 11 illustrates when stakeholder counts are well spread across the topics, their $P$ values are low irregardless of high or low counts, which is the desired behaviour since stakeholder with such behaviours fall under either `Normal' or `Irrelevant' (Tab.~\ref{tab:freq_table}). Alternatively, `Significant' cases with strong single-peak distribution dominating over the others (Indices 4, 5, 8, 9) show significantly higher $P$ values, where completely unique stakeholders tend to reach maximum possible $P$. It is noteworthy to add that the difference between the final $Rel$ score between cases 8 and 9 is important since the $R$ scores are able to dampen case 8's score due to its low count value which was set to the common median (3) for illustration. This also occurs between cases 4 and 5. Lastly, those with double-peak distribution (Indices 6 and 7) have a mid-low $P$ value, where the duo peaks' $P$ are significantly higher than that of the topic with the low-value count, yet not as high as what would occur in a single-peak. The duo-peaks were penalized with the number of similarly `Significant' topics within themselves. This leads to a lower overall $Rel$ score.

\newpage

\begin{figure*}[ht]
  \centering
  \includegraphics[width=\linewidth]{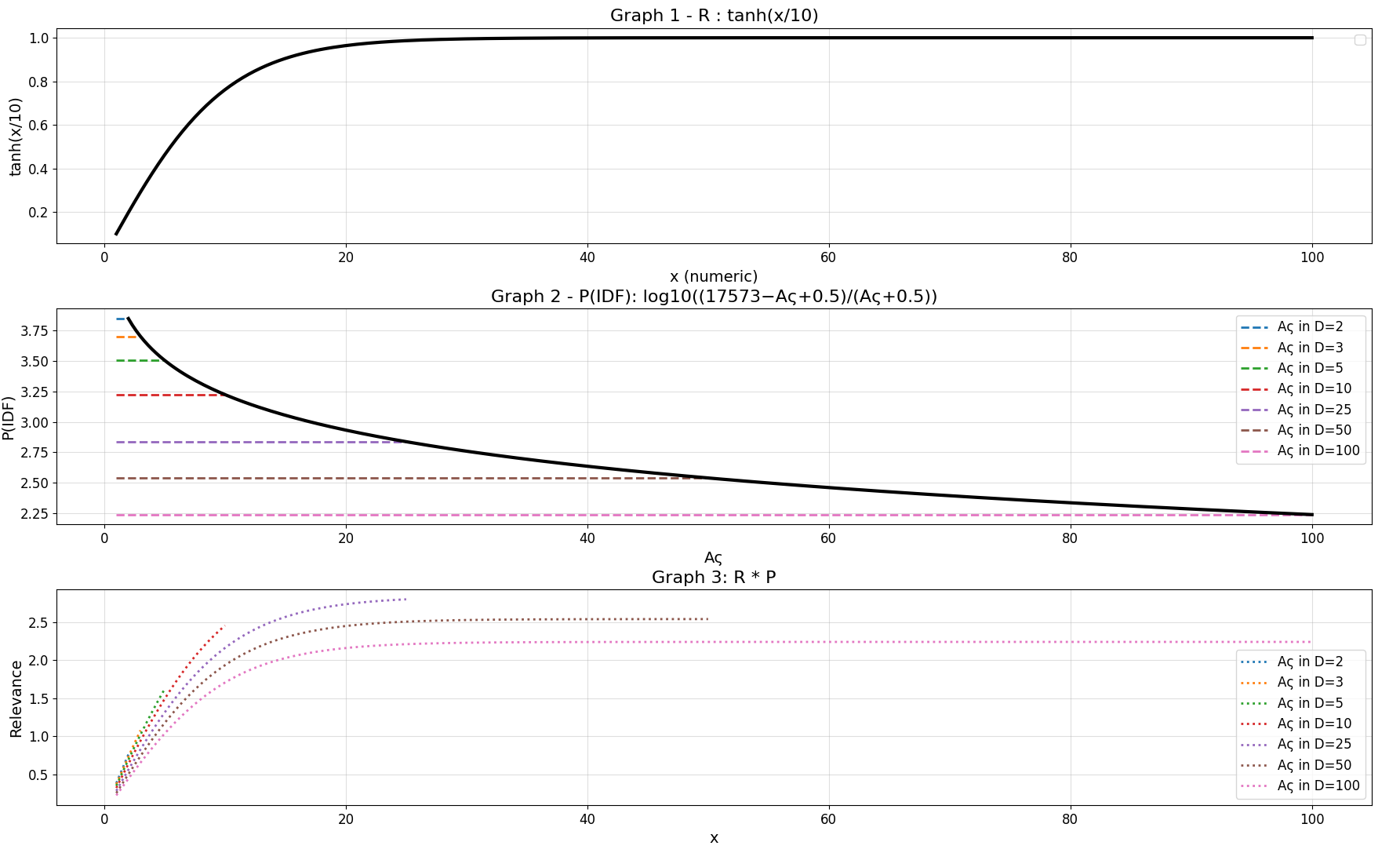}
  \caption{Representation of Penalty$_{IDF}$ Behaviour}
  \label{fig:pv1crisis}
\end{figure*}

Similarly, we will examine the behaviour of $P_{IDF}$. Graph 1 of Figure~\ref{fig:pv1crisis} also shows $R$ for reference. Graph 2 plots unnormalized $P_{IDF}$ value to show how it changes as $|A_\varsigma \in D|$ increases (Eqn.~\ref{eqn:P1}). It should be noted that indices 1, 3, 4, 7, 9 and 11 (Tab.~\ref{tab:crisispv2sit}) cannot be differentiated between topics (i.e. A, B and C) since there is no differentiation between topics while using $P_{IDF}$ which makes it inferior to $P$. Despite so, it can be observed from Graph 2 that as more articles uses the same stakeholder, $P_{IDF}$ becomes smaller, mirroring the way inverse document frequencies work. 

Graph 3 further documents the interaction of $R$ and $P_{IDF}$, where every line shows how the differing counts of stakeholders across topic ($|A_\varsigma \in D|$) interact when $P_{IDF}$ multiplies $R$ which rewards within a topic ($count(\varsigma_d)$). It is observed that the relevance score of `Irrelevant' cases (Tab.~\ref{tab:freq_table}) will still be highly rewarded beyond 1 once $count(\varsigma_d)$ goes beyond 5 regardless of $|A_\varsigma \in D|$, for instance having 5 out of 100 of any stakeholder $\varsigma$ allows a $Rel$ of around 1, which is not ideal despite its proportionally low count value. This flaw from $P_{IDF}$ shows that proportion and distribution should be essential, which is accounted for within $P$.

\newpage

\section{Repeating Stakeholder Counts}~\label{appn:stakeC}

\begin{figure*}[!h]
  \centering
  \includegraphics[width=\linewidth]{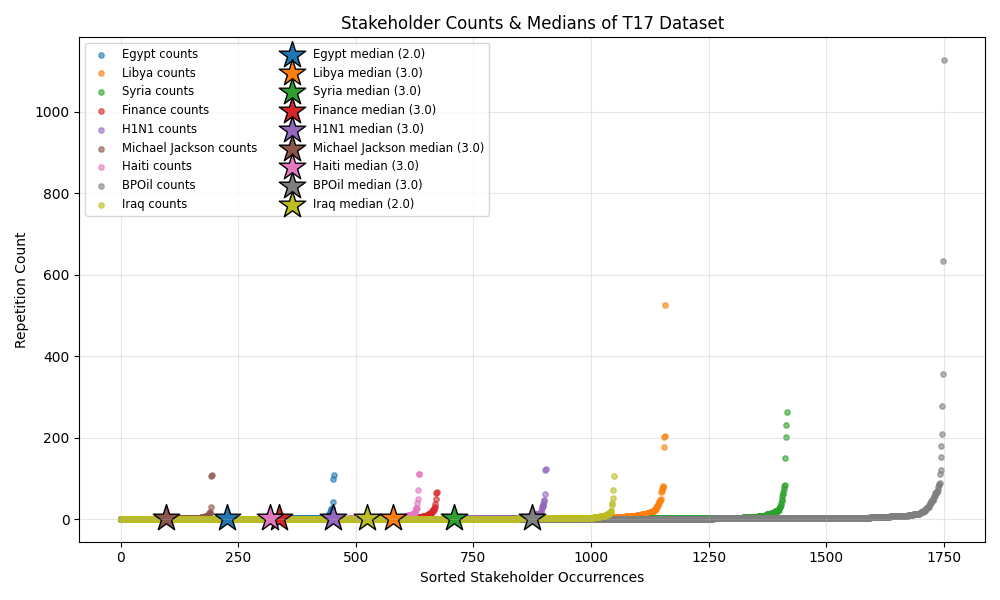}
  \caption{Stakeholder counts across topics within T17 Dataset}
  \label{fig:staket17}
  \vspace{4em}
  \centering
  \includegraphics[width=\linewidth]{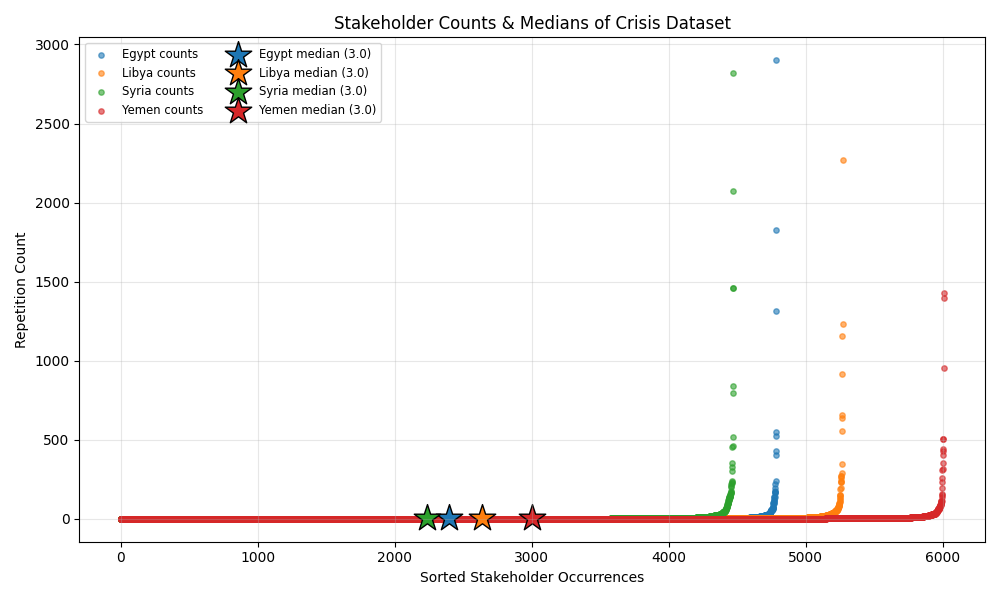}
  \caption{Stakeholder counts across topics within Crisis Dataset}
  \label{fig:stakecrisis}
\end{figure*}

We recorded the counts for each stakeholder across all topics in our dataset. There are occasional outliers where the entity involved may be mentioned more than a hundred times and may even reach the thousands, but typically, a count of ten or less is normal across all stakeholders existing in all topics. From Figures~\ref{fig:staket17} and \ref{fig:stakecrisis}, we can observe that the individual median of each topic regardless of the dataset is around three. This means that within their own topics, most stakeholders are mentioned at least in three different articles.

\newpage
\clearpage
\section{Full Set of Experiment Results}~\label{appn:expf}

\begin{table}[H] %v2 crisis p2 then %v1 crisis p2
\caption{Crisis Results-SUnSET with $P$ for $Rel$}
    \centering
    \begin{tabular}{|l l|l l l|l l l|}
    \hline
        \multicolumn{2}{|c|}{} & \multicolumn{3}{c|}{TextRank+$Rel$} & \multicolumn{3}{c|}{TextRank} \\ 
        beta & EM & AR-1 & AR-2 & Date-F1 & AR-1 & AR-2 & Date-F1  \\ \hline
        0 & 0 & 0.114358 & 0.041683 & 0.354251  & 0.114358 & 0.041683 & 0.354251  \\ 
        0.1 & 0 & 0.113652 & 0.039741 & 0.35494 & 0.112811 & 0.039234 & 0.353606  \\ 
        0.2 & 0 & 0.116541 & 0.041465 & 0.360721 & 0.113745 & 0.039872 & 0.354645  \\
        0.3 & 0 & 0.118841 & 0.043264 & 0.367969 & 0.11459 & 0.040077 & 0.358176  \\
        0.4 & 0 & 0.119222 & 0.043375 & 0.367698 & 0.113476 & 0.039522 & 0.357489  \\
        0.5 & 0 & 0.122711 & 0.04451 & 0.377522  & 0.116135 & 0.039776 & 0.366623  \\
        0.6 & 0 & 0.12393 & 0.04462 & 0.381933 & 0.116748 & 0.040169 & 0.368  \\ 
        0.7 & 0 & 0.125512 & 0.0456 & 0.384202 & 0.116847 & 0.040308 & 0.368  \\
        0.8 & 0 & 0.127489 & 0.047185 & 0.383934 & 0.116723 & 0.040229 & 0.364686  \\ 
        0.9 & 0 & 0.12894 & 0.047628 & 0.387541 & 0.116548 & 0.040077 & 0.364686  \\
        1 & 0 & 0.129269 & 0.047488 & 0.388972  & 0.116641 & 0.04008 & 0.364686  \\ \hline
        0 & 1 & 0.119383 & 0.044027 & 0.364554 & 0.119383 & 0.044027 & 0.364554  \\ 
        0.1 & 1 & 0.11499 & 0.040711 & 0.358591 & 0.114242 & 0.040103 & 0.357267  \\
        0.2 & 1 & 0.117851 & 0.04203 & 0.363621 & 0.115888 & 0.041408 & 0.361055  \\
        0.3 & 1 & 0.119364 & 0.042948 & 0.368201 & 0.115872 & 0.041243 & 0.361055  \\
        0.4 & 1 & 0.119155 & 0.042993 & 0.37158 & 0.115795 & 0.04119 & 0.360607  \\
        0.5 & 1 & 0.120324 & 0.043291 & 0.374345 & 0.116859 & 0.040401 & 0.366181  \\
        0.6 & 1 & 0.123812 & 0.044156 & 0.385318 & 0.116893 & 0.040489 & 0.366181  \\
        0.7 & 1 & 0.124445 & 0.044881 & 0.383322 & 0.117104 & 0.040751 & 0.366181  \\
        0.8 & 1 & 0.125976 & 0.04591 & 0.383054 & 0.117077 & 0.040633 & 0.366181  \\
        0.9 & 1 & 0.127911 & 0.046209 & 0.386313 & 0.116135 & 0.039776 & 0.366623  \\
        1 & 1 & 0.124948 & 0.045573 & 0.374559 & 0.114044 & 0.039693 & 0.366867  \\ \hline
        0 & 2 & 0.114743 & 0.041232 & 0.358151 & 0.114743 & 0.041232 & 0.358151  \\
        0.1 & 2 & 0.114549 & 0.039838 & 0.365324 & 0.113659 & 0.039767 & 0.361014  \\
        0.2 & 2 & 0.115156 & 0.039829 & 0.367784 & 0.113425 & 0.039519 & 0.361014  \\
        0.3 & 2 & 0.118806 & 0.041571 & 0.371988 & 0.114599 & 0.040124 & 0.361038  \\
        0.4 & 2 & 0.120898 & 0.04296 & 0.37482 & 0.115251 & 0.040805 & 0.363499  \\
        0.5 & 2 & 0.121954 & 0.043179 & 0.377068 & 0.115556 & 0.040812 & 0.363499  \\
        0.6 & 2 & 0.121427 & 0.04293 & 0.379509 & 0.116569 & 0.040968 & 0.365772  \\
        0.7 & 2 & 0.120243 & 0.042156 & 0.374532 & 0.1158 & 0.040556 & 0.364149  \\
        0.8 & 2 & 0.120882 & 0.042673 & 0.375772 & 0.115812 & 0.040559 & 0.364149  \\
        0.9 & 2 & 0.122365 & 0.043452 & 0.375772 & 0.115122 & 0.04015 & 0.365345  \\
        1 & 2 & 0.12309 & 0.043711 & 0.375196 & 0.115122 & 0.040148 & 0.365345  \\ \hline
    \end{tabular}
\end{table}

\begin{table}[ht] %v2 t17 p2 -->%v1 t17 p2
\caption{T17 Results-SUnSET with $P$ for $Rel$}
    \centering
    \begin{tabular}{|l l|l l l|l l l|}
    \hline
        \multicolumn{2}{|c|}{} & \multicolumn{3}{c|}{TextRank+$Rel$} & \multicolumn{3}{c|}{TextRank} \\ 
        beta & EM & AR-1 & AR-2 & Date-F1 & AR-1 & AR-2 & Date-F1  \\ \hline
        0 & 0 & 0.127151 & 0.038522 & 0.551891 & 0.127151 & 0.038522 & 0.551891  \\
        0.1 & 0 & 0.133391 & 0.042983 & 0.567552 & 0.131371 & 0.042254 & 0.554819  \\
        0.2 & 0 & 0.132674 & 0.042788 & 0.566986 & 0.128913 & 0.040551 & 0.550866  \\
        0.3 & 0 & 0.13343 & 0.04304 & 0.569759 & 0.128958 & 0.040486 & 0.556219  \\
        0.4 & 0 & 0.13434 & 0.043684 & 0.571095 & 0.128798 & 0.040446 & 0.556219  \\ 
        0.5 & 0 & 0.136108 & 0.044275 & 0.571878 & 0.129042 & 0.040489 & 0.556219  \\
        0.6 & 0 & 0.135763 & 0.044037 & 0.571878 & 0.12782 & 0.03994 & 0.55434  \\
        0.7 & 0 & 0.13604 & 0.044052 & 0.569624 & 0.128425 & 0.040278 & 0.554834  \\
        0.8 & 0 & 0.13582 & 0.043954 & 0.571009 & 0.128086 & 0.040158 & 0.554834  \\
        0.9 & 0 & 0.134498 & 0.043165 & 0.574814 & 0.127756 & 0.03997 & 0.554834  \\ 
        1 & 0 & 0.134713 & 0.043138 & 0.574814 & 0.127738 & 0.039948 & 0.554834  \\ \hline
        0 & 1 & 0.128352 & 0.040235 & 0.553264 & 0.127691 & 0.039799 & 0.551384  \\
        0.1 & 1 & 0.134273 & 0.043627 & 0.56971 & 0.131719 & 0.042692 & 0.553908  \\
        0.2 & 1 & 0.133477 & 0.043096 & 0.571774 & 0.129177 & 0.041106 & 0.550931  \\ 
        0.3 & 1 & 0.134157 & 0.043373 & 0.571774 & 0.129145 & 0.041125 & 0.55464  \\ 
        0.4 & 1 & 0.134424 & 0.043731 & 0.571774 & 0.128781 & 0.040995 & 0.55464  \\
        0.5 & 1 & 0.136009 & 0.044221 & 0.572557 & 0.128841 & 0.040975 & 0.55464  \\
        0.6 & 1 & 0.136132 & 0.044262 & 0.572557 & 0.127743 & 0.040477 & 0.549568  \\
        0.7 & 1 & 0.136607 & 0.044352 & 0.571172 & 0.128484 & 0.041047 & 0.550062 \\
        0.8 & 1 & 0.136573 & 0.044345 & 0.573073 & 0.128222 & 0.040971 & 0.550062  \\ 
        0.9 & 1 & 0.136085 & 0.043953 & 0.576362 & 0.128222 & 0.040947 & 0.550062  \\ 
        1 & 1 & 0.136176 & 0.043886 & 0.576362 & 0.128222 & 0.040947 & 0.550062  \\ \hline
        0 & 2 & 0.127571 & 0.04065 & 0.558906 & 0.127571 & 0.04065 & 0.558906  \\
        0.1 & 2 & 0.134279 & 0.044326 & 0.567725 & 0.127598 & 0.041011 & 0.548946  \\
        0.2 & 2 & 0.132227 & 0.04378 & 0.565846& 0.126087 & 0.039899 & 0.548946  \\ 
        0.3 & 2 & 0.13283 & 0.043909 & 0.57041 & 0.126327 & 0.03999 & 0.55172  \\
        0.4 & 2 & 0.134523 & 0.04468 & 0.57041 & 0.127085 & 0.040477 & 0.55172  \\ 
        0.5 & 2 & 0.133141 & 0.043959 & 0.57041 & 0.127483 & 0.040714 & 0.55172  \\
        0.6 & 2 & 0.133539 & 0.044212 & 0.57041 & 0.127142 & 0.040493 & 0.55172  \\ 
        0.7 & 2 & 0.133565 & 0.043979 & 0.57041 & 0.127077 & 0.040308 & 0.55172  \\ 
        0.8 & 2 & 0.133548 & 0.043975 & 0.571507 & 0.126337 & 0.040098 & 0.55172  \\
        0.9 & 2 & 0.133639 & 0.04401 & 0.571507 & 0.124933 & 0.038988 & 0.54984  \\ 
        1 & 2 & 0.13477 & 0.044337 & 0.573387 & 0.124747 & 0.039042 & 0.54984  \\ \hline
    \end{tabular}
\end{table}

\begin{table}[ht] %v2 crisis p1--> %v1 crisis p1
    \centering
    \caption{Crisis Results-SUnSET with $P_{IDF}$ for $Rel$}
    \begin{tabular}{|l l|l l l|l l l|}
    \hline
        \multicolumn{2}{|c|}{} & \multicolumn{3}{c|}{TextRank+$Rel$} & \multicolumn{3}{c|}{TextRank} \\ 
        beta & EM & AR-1 & AR-2 & Date-F1 & AR-1 & AR-2 & Date-F1  \\ \hline
        0 & 0 & 0.114358 & 0.041683 & 0.354251 & 0.114358 & 0.041683 & 0.354251  \\
        0.1 & 0 & 0.119213 & 0.042284 & 0.364802& 0.11779 & 0.040881 & 0.363209   \\ 
        0.2 & 0 & 0.11937 & 0.042153 & 0.365167  & 0.117337 & 0.040573 & 0.363209 \\
        0.3 & 0 & 0.119812 & 0.042737 & 0.369653 & 0.117364 & 0.040541 & 0.363942  \\ 
        0.4 & 0 & 0.118798 & 0.042092 & 0.370735  & 0.118033 & 0.041105 & 0.367189  \\ 
        0.5 & 0 & 0.119503 & 0.043181 & 0.375763 & 0.118011 & 0.041095 & 0.367189   \\
        0.6 & 0 & 0.118468 & 0.042296 & 0.37138 & 0.118011 & 0.041095 & 0.367189 \\ 
        0.7 & 0 & 0.118511 & 0.042324 & 0.370646 & 0.118011 & 0.041095 & 0.367189   \\
        0.8 & 0 & 0.118789 & 0.042531 & 0.369889  & 0.118011 & 0.041095 & 0.367189  \\ 
        0.9 & 0 & 0.119232 & 0.043089 & 0.373136 & 0.118011 & 0.041095 & 0.367189   \\
        1 & 0 & 0.118836 & 0.042856 & 0.373136 & 0.118011 & 0.041095 & 0.367189  \\ \hline
        0 & 1 & 0.119383 & 0.044027 & 0.364554 & 0.119383 & 0.044027 & 0.364554 \\
        0.1 & 1 & 0.116948 & 0.041352 & 0.368236 & 0.11742 & 0.041117 & 0.3668  \\ 
        0.2 & 1 & 0.114585 & 0.040833 & 0.357094 & 0.117598 & 0.041219 & 0.3668   \\ 
        0.3 & 1 & 0.114909 & 0.041108 & 0.362224   & 0.11765 & 0.041245 & 0.367534\\ 
        0.4 & 1 & 0.116967 & 0.041394 & 0.371358  & 0.117646 & 0.041252 & 0.367534  \\ 
        0.5 & 1 & 0.117696 & 0.042491 & 0.373769 & 0.117646 & 0.041252 & 0.367534   \\ 
        0.6 & 1 & 0.117333 & 0.042068 & 0.37128 & 0.117646 & 0.041252 & 0.367534  \\ 
        0.7 & 1 & 0.119144 & 0.043011 & 0.374527 & 0.117646 & 0.041252 & 0.367534  \\ 
        0.8 & 1 & 0.118695 & 0.042807 & 0.373769 & 0.117646 & 0.041252 & 0.367534   \\ 
        0.9 & 1 & 0.118588 & 0.042762 & 0.373769  & 0.117646 & 0.041252 & 0.367534  \\ 
        1 & 1 & 0.117075 & 0.04204 & 0.372349 & 0.117646 & 0.041252 & 0.367534   \\ \hline
        0 & 2 & 0.114743 & 0.041232 & 0.358151  & 0.114743 & 0.041232 & 0.358151 \\  
        0.1 & 2 & 0.113404 & 0.039178 & 0.366255 & 0.112941 & 0.039047 & 0.363297   \\  
        0.2 & 2 & 0.112839 & 0.039618 & 0.360665  & 0.112568 & 0.03918 & 0.361919 \\ 
        0.3 & 2 & 0.112525 & 0.038941 & 0.361231 & 0.113128 & 0.039125 & 0.364192  \\  
        0.4 & 2 & 0.112189 & 0.038723 & 0.362651 & 0.113128 & 0.039093 & 0.364192  \\  
        0.5 & 2 & 0.113626 & 0.039906 & 0.364744 & 0.113128 & 0.039093 & 0.364192  \\ 
        0.6 & 2 & 0.112348 & 0.039172 & 0.363323 & 0.112722 & 0.03876 & 0.364192  \\  
        0.7 & 2 & 0.112828 & 0.03947 & 0.364744  & 0.11269 & 0.038761 & 0.364192  \\ 
        0.8 & 2 & 0.112622 & 0.039416 & 0.364281 & 0.11269 & 0.038761 & 0.364192   \\
        0.9 & 2 & 0.112912 & 0.039573 & 0.364281  & 0.11269 & 0.038761 & 0.364192  \\ 
        1 & 2 & 0.113113 & 0.039672 & 0.363477 & 0.11269 & 0.038761 & 0.364192   \\ \hline
    \end{tabular}
\end{table}

\begin{table}[ht] %v2 t17 p1-->%v1 t17 p1
    \centering
    \caption{T17 Results-SUnSET with $P_{IDF}$ for $Rel$}
    \begin{tabular}{|l l|l l l|l l l|}
    \hline
        \multicolumn{2}{|c|}{} & \multicolumn{3}{c|}{TextRank+$Rel$} & \multicolumn{3}{c|}{TextRank} \\ 
        beta & EM & AR-1 & AR-2 & Date-F1 & AR-1 & AR-2 & Date-F1  \\ \hline
        0 & 0 & 0.127151 & 0.038522 & 0.551891 & 0.127151 & 0.038522 & 0.551891  \\ 
        0.1 & 0 & 0.133361 & 0.042997 & 0.571505 & 0.128408 & 0.040082 & 0.55271   \\  
        0.2 & 0 & 0.133766 & 0.042873 & 0.575374 & 0.128416 & 0.039956 & 0.554628 \\  
        0.3 & 0 & 0.134439 & 0.042803 & 0.575513  & 0.128083 & 0.039733 & 0.556272 \\  
        0.4 & 0 & 0.135246 & 0.043226 & 0.575513 & 0.128149 & 0.039836 & 0.556272  \\  
        0.5 & 0 & 0.135214 & 0.043271 & 0.574128  & 0.128149 & 0.039836 & 0.556272   \\  
        0.6 & 0 & 0.135432 & 0.043603 & 0.573612  & 0.128149 & 0.039836 & 0.556272 \\  
        0.7 & 0 & 0.135341 & 0.043572 & 0.575207 & 0.128149 & 0.039836 & 0.556272   \\  
        0.8 & 0 & 0.134963 & 0.043413 & 0.575658  & 0.12811 & 0.039807 & 0.556272 \\  
        0.9 & 0 & 0.13518 & 0.043546 & 0.574529 & 0.12811 & 0.039807 & 0.556272  \\ 
        1 & 0 & 0.133889 & 0.042507 & 0.57265 & 0.12811 & 0.039807 & 0.556272  \\ \hline
        0 & 1 & 0.128352 & 0.040235 & 0.553264 & 0.127691 & 0.039799 & 0.551384   \\ 
        0.1 & 1 & 0.137057 & 0.044288 & 0.574131  & 0.129329 & 0.040677 & 0.545189 \\  
        0.2 & 1 & 0.136771 & 0.043724 & 0.576053 & 0.129147 & 0.040417 & 0.548139 \\  
        0.3 & 1 & 0.13739 & 0.043633 & 0.576029 & 0.128679 & 0.040163 & 0.548139 \\  
        0.4 & 1 & 0.137267 & 0.043609 & 0.576029 & 0.129101 & 0.040205 & 0.549784  \\ 
        0.5 & 1 & 0.137116 & 0.043592 & 0.574128  & 0.129101 & 0.040205 & 0.549784  \\  
        0.6 & 1 & 0.137352 & 0.043926 & 0.573612 & 0.129101 & 0.040205 & 0.549784  \\ 
        0.7 & 1 & 0.136805 & 0.043857 & 0.573612 & 0.129101 & 0.040205 & 0.549784  \\  
        0.8 & 1 & 0.136675 & 0.043737 & 0.572515 & 0.129101 & 0.040205 & 0.549784   \\ 
        0.9 & 1 & 0.136709 & 0.043784 & 0.571386 & 0.129101 & 0.040205 & 0.549784   \\  
        1 & 1 & 0.135355 & 0.04276 & 0.569507 & 0.129101 & 0.040205 & 0.549784   \\ \hline
        0 & 2 & 0.127571 & 0.04065 & 0.558906   & 0.127571 & 0.04065 & 0.558906 \\  
        0.1 & 2 & 0.134797 & 0.043389 & 0.574589 & 0.126857 & 0.040582 & 0.551637   \\  
        0.2 & 2 & 0.134322 & 0.043568 & 0.57271 & 0.125526 & 0.039409 & 0.553047  \\ 
        0.3 & 2 & 0.134659 & 0.043783 & 0.573226 & 0.125708 & 0.039558 & 0.553047 \\ 
        0.4 & 2 & 0.134008 & 0.042938 & 0.57271 & 0.125539 & 0.039507 & 0.552531  \\  
        0.5 & 2 & 0.13373 & 0.043024 & 0.57271 & 0.12594 & 0.039747 & 0.552531 \\ 
        0.6 & 2 & 0.13373 & 0.043039 & 0.57271 & 0.12594 & 0.039747 & 0.552531  \\  
        0.7 & 2 & 0.134293 & 0.043185 & 0.57271 & 0.12594 & 0.039747 & 0.552531  \\  
        0.8 & 2 & 0.134293 & 0.043185 & 0.57271  & 0.12594 & 0.039747 & 0.552531\\  
        0.9 & 2 & 0.134407 & 0.043212 & 0.57271  & 0.12594 & 0.039747 & 0.552531 \\ 
        1 & 2 & 0.134075 & 0.043106 & 0.57271 & 0.12594 & 0.039747 & 0.552531  \\ \hline
    \end{tabular}
\end{table}

% \newpage
% \clearpage
% \section{Textual Examples of Timeline Generation}
% \tea{Basura: give example: If you use stakeholder you get this text, if you don't you get this, textual examples.}

\clearpage

\section{Example Timelines Generated from Experiments}~\label{appn:example}

\begin{figure*}[!ht]
  \centering
  % Top image
  \includegraphics[width=0.75\linewidth]{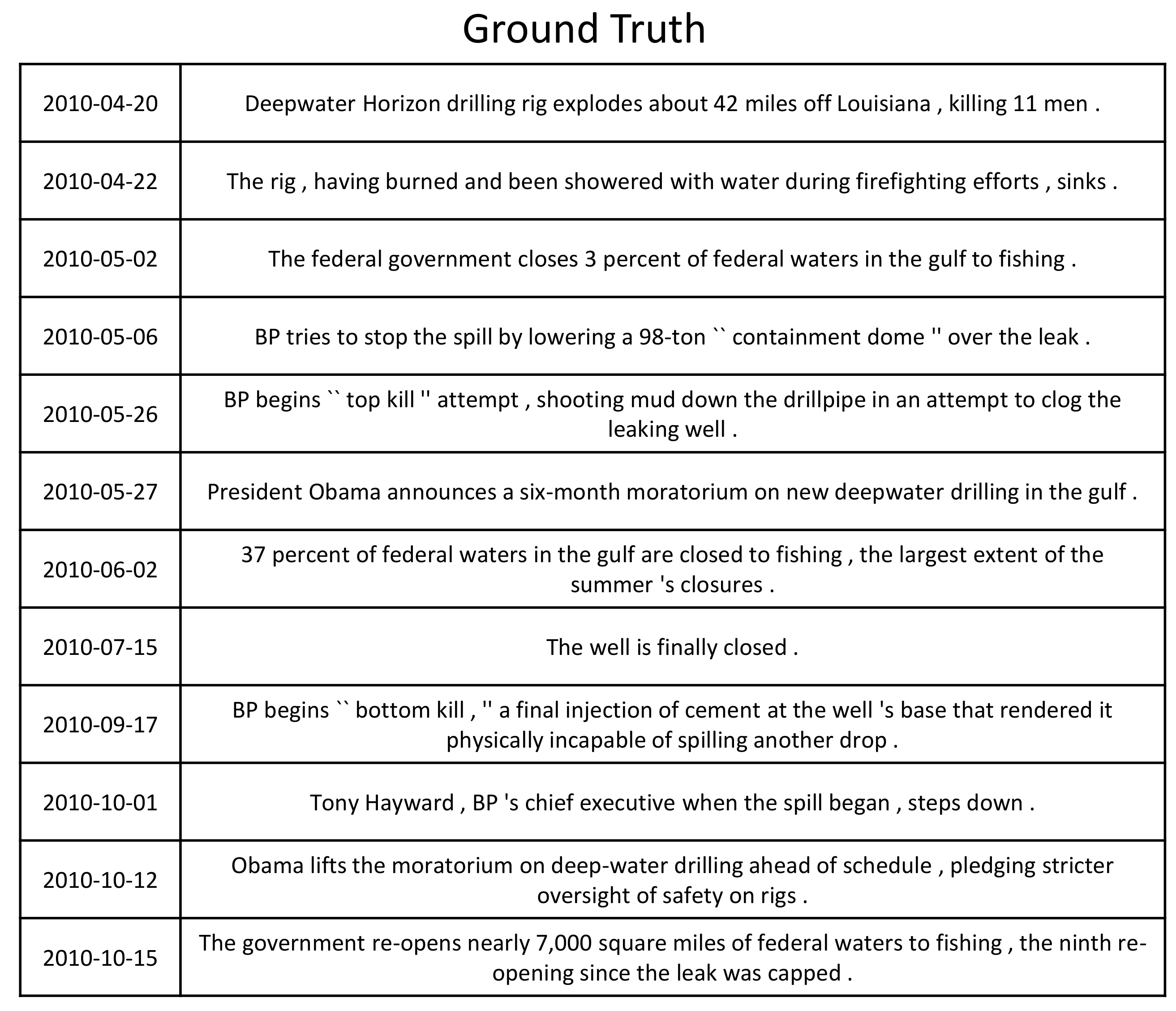}
  \caption{Example of Ground Truth Timeline of T17 BP-Oil}
  \label{fig:top}
  
  \vspace{0.5em} % Adds space between top and bottom row

  % Bottom two images side-by-side stretched beyond margins
  \begin{adjustwidth*}{-2.3cm}{-2.3cm} % adjust the values as needed
    \begin{subfigure}[b]{0.5\linewidth}
      \includegraphics[width=\linewidth]{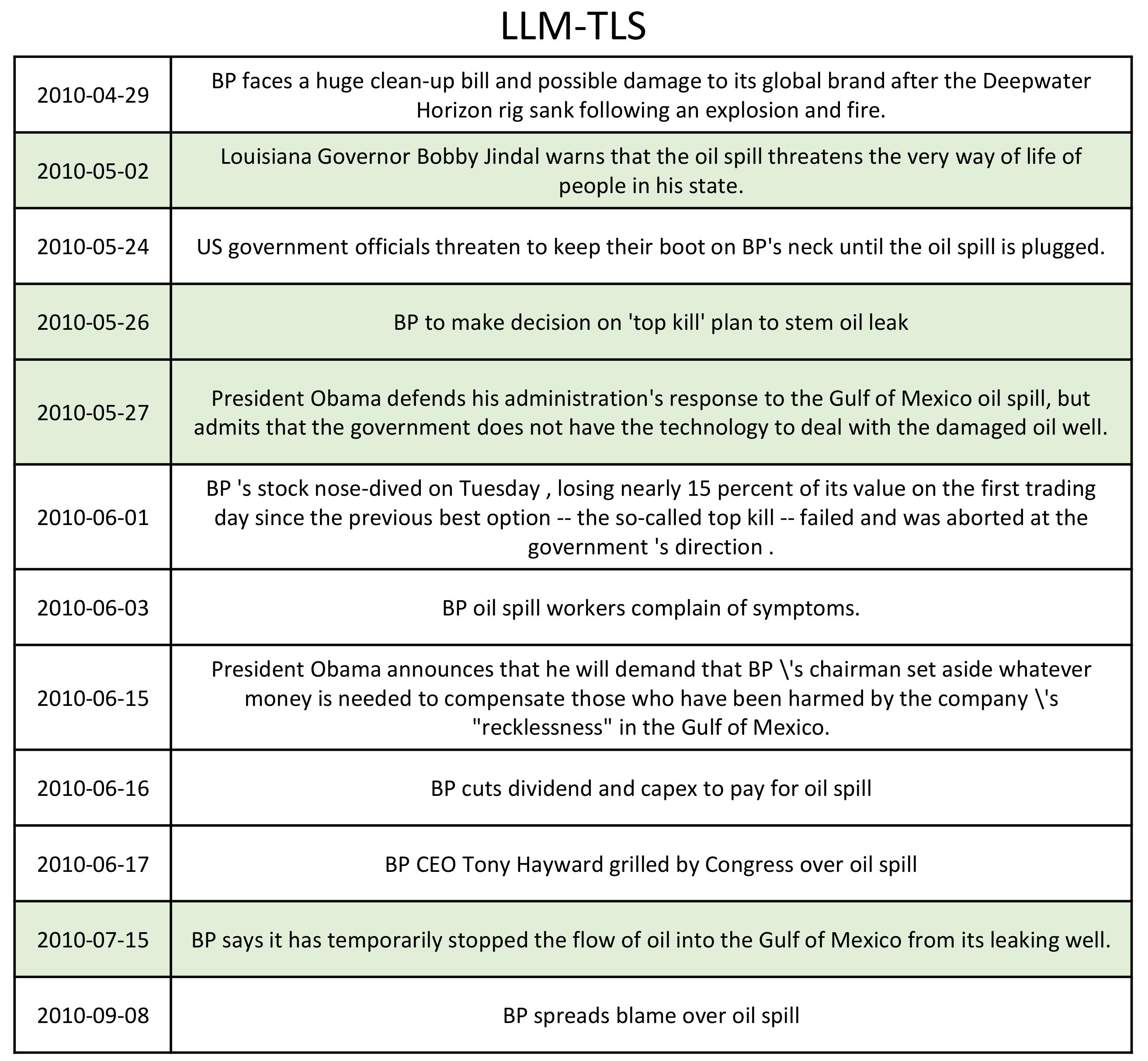}
      \caption{LLM-TLS Generated Timeline}
      \label{fig:bottomleft}
    \end{subfigure}
    \begin{subfigure}[b]{0.5\linewidth}
      \includegraphics[width=\linewidth]{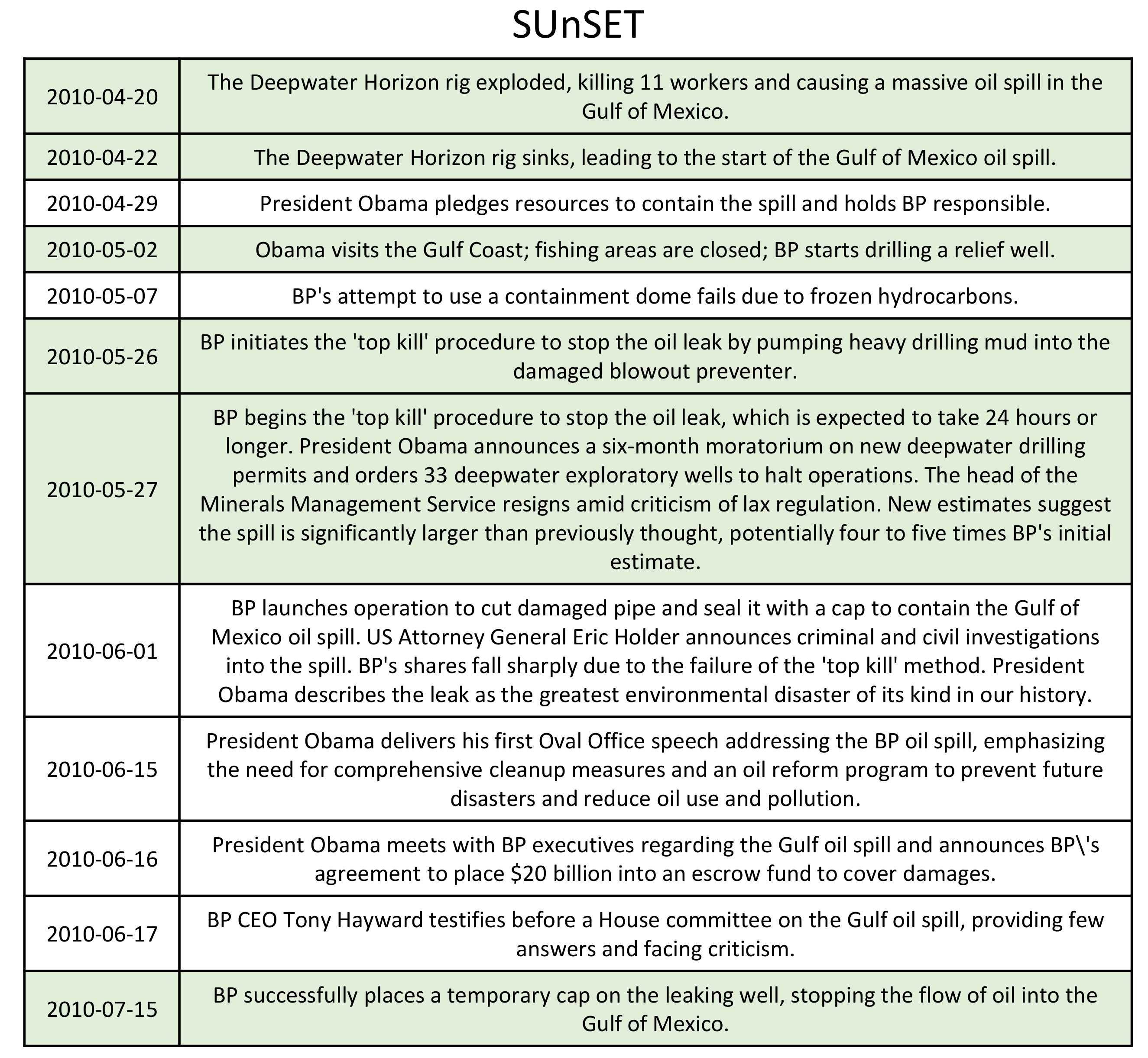}
      \caption{SUnSET Generated Timeline}
      \label{fig:bottomright}
    \end{subfigure}
  \end{adjustwidth*}

  \caption{LLM-TLS \textit{(a)} versus SUnSET \textit{(b)} in generating example timeline of T17 BP-Oil (Fig.~\ref{fig:top}). Text with green highlight indicates event which aligns better with Ground Truth}
  \label{fig:composite}
\end{figure*}

% First page: Top image alone
\begin{figure}[ht]
  \centering
  \includegraphics[width=\linewidth]{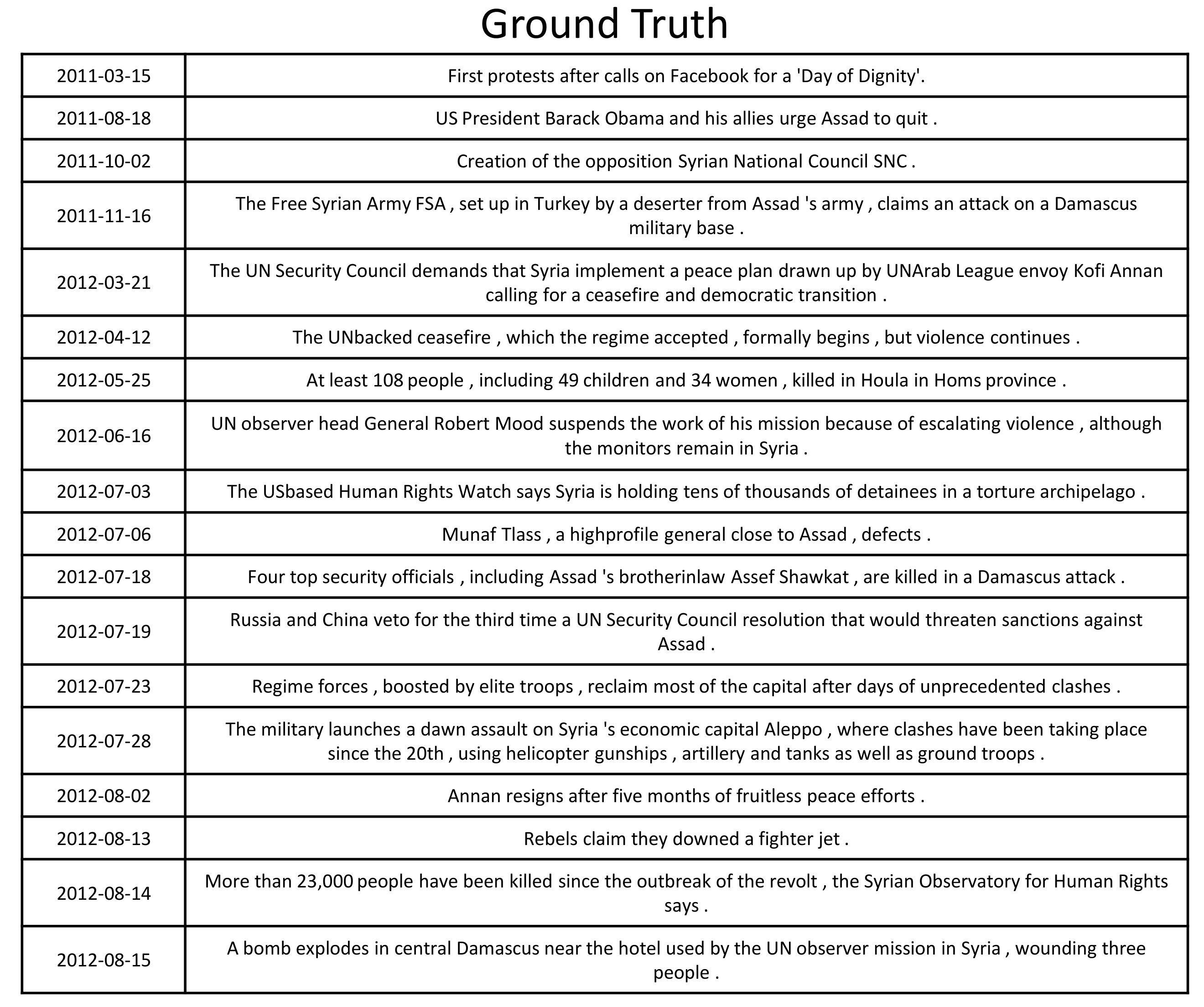}
  \caption{Example of Ground Truth Timeline of Crisis Syria}
  \label{fig:topb}
\end{figure}

\clearpage % Force the rest onto the next page

% Second page: stacked images
\begin{figure}[ht]
  \centering
  \includegraphics[width=0.8\linewidth]{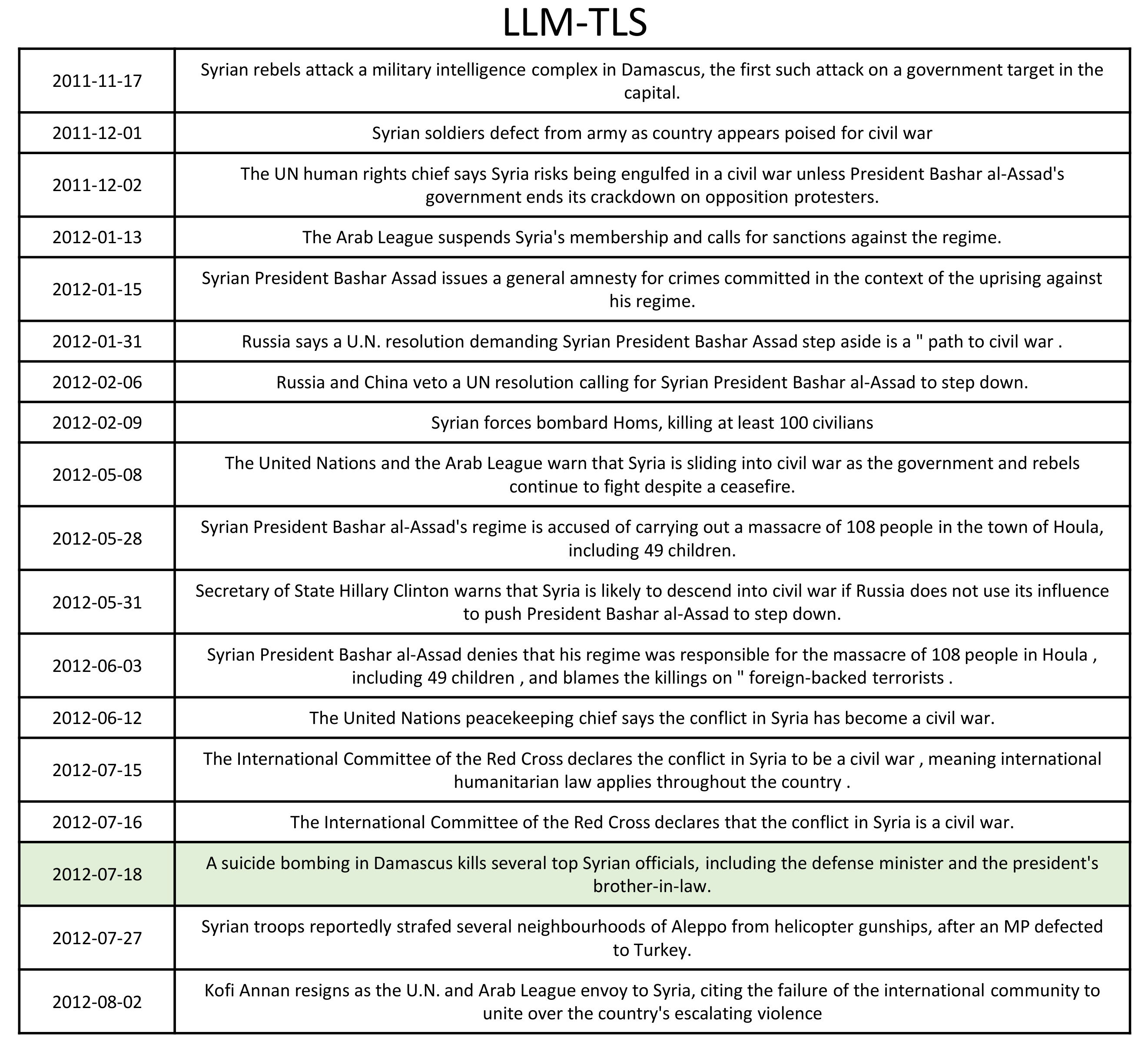}
  \caption{LLM-TLS Generated Timeline of Crisis Syria (Fig.~\ref{fig:topb})}
  \label{fig:stackedA}
  \centering
  \includegraphics[width=0.8\linewidth]{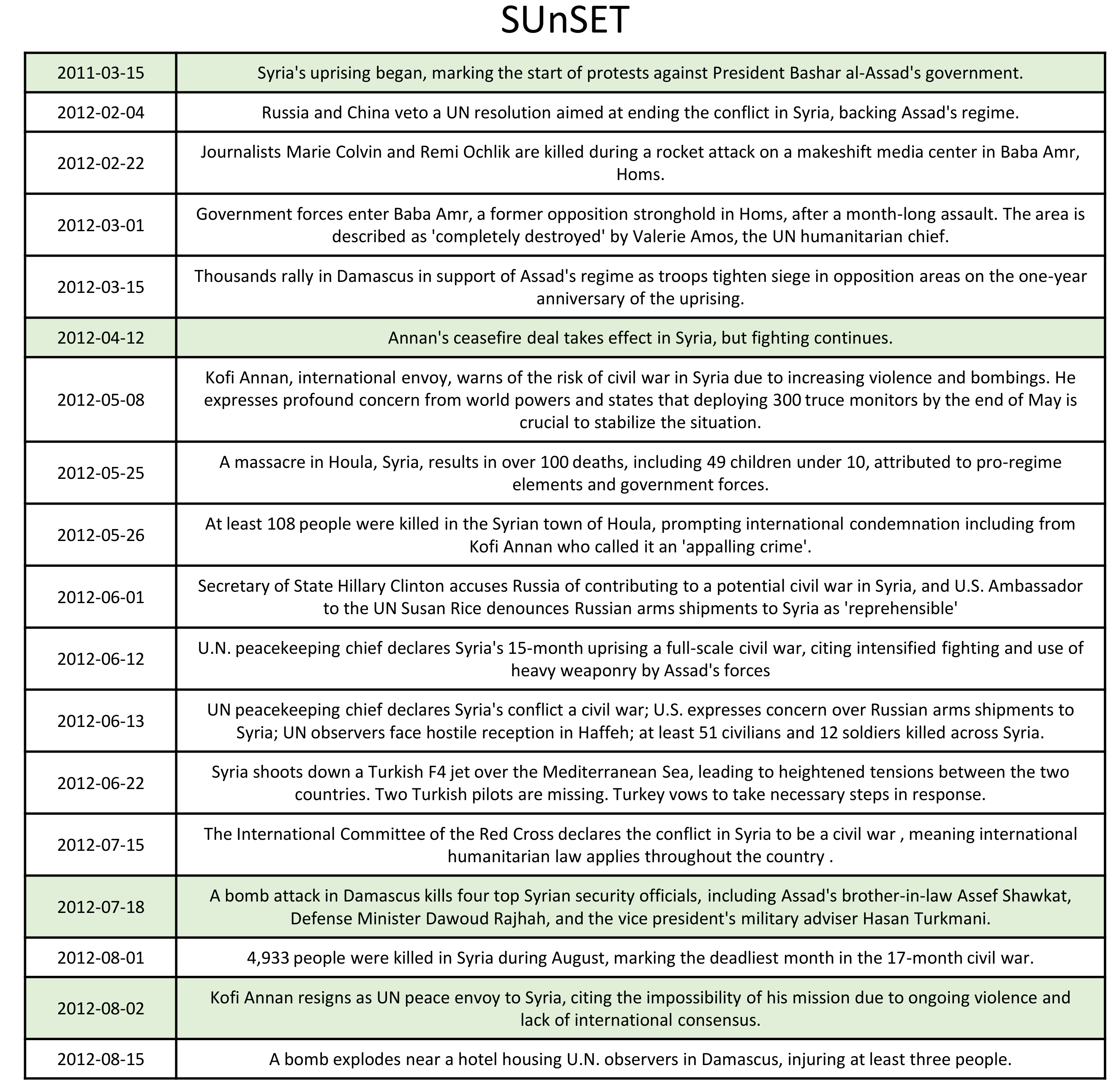}
  \caption{SUnSET Generated Timeline of Crisis Syria (Fig.~\ref{fig:topb})}
  \label{fig:stackedB}
\end{figure}
 % This imports the appendix.tex file

\end{document}